\documentclass[aps,pra,reprint,floatfix, superscriptaddress]{revtex4-2}
\usepackage{graphicx}
\usepackage{dcolumn}
\usepackage{bm}
\usepackage{hyperref}
\usepackage{natbib}
\usepackage{amsmath}
\usepackage{xcolor}
\usepackage{amssymb} 
\usepackage{braket} 
\usepackage{float}
\usepackage{comment}
\usepackage{dblfloatfix}

\begin{document}

\title{Pseudospectral method for solving PDEs using Matrix Product States}

\author{Jorge Gidi}
\email{jorgegidi@udec.cl}
\affiliation{Institute of Fundamental Physics IFF-CSIC, Calle Serrano 113b, Madrid 28006, Spain}
\affiliation{Millennium Institute for Research in Optics and Departamento de Física, Facultad de Ciencias F\'isicas y Matem\'aticas, Universidad de Concepci\'on, Casilla 160-C, Concepci\'on, Chile}

\author{Paula García-Molina}
\email{paula.garcia@iff.csic.es}
\affiliation{Institute of Fundamental Physics IFF-CSIC, Calle Serrano 113b, Madrid 28006, Spain}
\author{Luca Tagliacozzo}
\affiliation{Institute of Fundamental Physics IFF-CSIC, Calle Serrano 113b, Madrid 28006, Spain}
\author{Juan José García-Ripoll}
\affiliation{Institute of Fundamental Physics IFF-CSIC, Calle Serrano 113b, Madrid 28006, Spain}

\begin{abstract}
  This research focuses on solving time-dependent partial differential equations (PDEs), in particular the time-dependent Schrödinger equation, using matrix product states (MPS). We propose an extension of Hermite Distributed Approximating Functionals (HDAF) to MPS, a highly accurate pseudospectral method for approximating functions of derivatives. Integrating HDAF into an MPS finite precision algebra, we test four types of quantum-inspired algorithms for time evolution: explicit Runge-Kutta methods, Crank-Nicolson method, explicitly restarted Arnoldi iteration and split-step.
  The benchmark problem is the expansion of a particle in a quantum quench, characterized by a rapid increase in space requirements, where HDAF surpasses traditional finite difference methods in accuracy with a comparable cost.
  Moreover, the efficient HDAF approximation to the free propagator avoids the need for Fourier transforms in split-step methods, significantly enhancing their performance with an improved balance in cost and accuracy. 
  Both approaches exhibit similar error scaling and run times compared to FFT vector methods; however, MPS offer an exponential advantage in memory, overcoming vector limitations to enable larger discretizations and expansions. Finally, the MPS HDAF split-step method successfully reproduces the physical behavior of a particle expansion in a double-well potential, demonstrating viability for actual research scenarios.
\end{abstract}

\maketitle

\section{Introduction}

Solving time-dependent partial differential equations (PDEs) is crucial across most fields in science and engineering.
In the quantum domain, the challenge is additionally plagued by exponential costs~\cite{Feynman1982}, both in the number of components of the system and in the unbounded domain of certain problems. An example of the latter is a particle expansion in a potential well~\cite{Bonvin_2024, RodaLlordes2024}.

The simulation of large quantum systems rapidly becomes untractable by traditional computational approaches due to their exponential scaling in complexity. Quantum computers have been posed as a promising tool to solve PDEs more efficiently~\cite{Leyton_Osborne_2008}, making use of the exponential compression that the amplitude encoding provides~\cite{GarciaRipoll2021, GarciaMolina2022, GonzalezConde2024} and tools such as the Quantum Fourier Transform (QFT)~\cite{Hales_2000}. However, there are still no scalable and fault-tolerant quantum computers where those algorithms may be applied~\cite{Chen_Cotler_Huang_Li_2023}, and it is also uncertain if quantum computers are required, especially for bandwidth-limited functions with limited entanglement~\cite{GarciaRipoll2021, Jobst2023}.

A current challenge is to develop alternative, quantum-inspired algorithms that profit from some of the exponential compressions available in quantum algorithms and encodings while enabling the execution in classical computers.
This challenge involves developing three tools: (i) an efficient encoding of the PDE solution, (ii) a corresponding encoding of the PDE operator itself, and (iii) an algorithm for time evolution.
In this work, we address these three objectives in novel ways.
We use the representation of matrix product states (MPS) for bandwidth-limited functions~\cite{Jobst2023}, also known in the mathematical community as quantized tensor trains (QTT)~\cite{Khoromskij2011}. Within this formulation, we develop an innovative encoding of differential operators, PDEs, and free-evolution propagators based on Hermite Distributed Approximating Functionals (HDAF)~\cite{Hoffman_Nayar_Sharafeddin_Kouri_1991, Hoffman_Kouri_1992, Hoffman_Arnold_Kouri_1992, Hoffman_Kouri_1995,hoffman_variational_1996, Bodmann_Hoffman_Kouri_Papadakis_2008}, to yield accurate matrix product operators (MPOs) with a low bond dimension. We present three families of time evolution algorithms using these operators: Global explicit and implicit methods, Arnoldi-based methods, and split-step methods. These new techniques are benchmarked against each other and alternative quantum-inspired methods based on finite differences~\cite{GarciaRipoll2021, GarciaMolina2023}. They are also compared to state-of-the-art spectral split-step methods in the standard vector representation~\cite{Weideman_Herbst_1986}.

As a benchmark, we study the expansion of a particle in a potential well. This computationally demanding scenario is highly relevant in optomechanics research~\cite{Levitodynamics, Bonvin_2024, RodaLlordes2024}. As the domain size increases dramatically, conventional computational methods suffer considerable strain. In this context, the qubit encoding in MPS/QTT may provide exponential data compression. However, the particle's acceleration induces chirping of the wavefunction, potentially increasing the bond dimension required beyond any computational benefit. For the limited cases where this problem can be analytically solved~\cite{GarciaRipoll_1999}, it constitutes an effective testbed to stress and validate numerical solvers.

When correctly tuned, the HDAF MPOs for differential operators are much more accurate than those based on finite differences and have a comparable cost. Also, the time evolution methods built upon these operators inherit an improvement in accuracy and efficiency.
Our best-performing time evolution method is the HDAF split-step, which leverages the efficient HDAF representation of the free propagator on a coordinate basis to avoid using Fourier Transforms.

For the particle expansion problem, quantum-inspired methods converge with subexponential time scaling, which is competitive with a vector implementation. Moreover, in the absence of chirping, our results are favorable to the HDAF split-step method using MPS, in contrast to vectors using the Fast Fourier Transform (FFT).

Quantum-inspired methods have been used to tackle quantum numerical analysis problems before, such as static~\cite{Lubasch2018, GarciaMolina2023} and time-dependent PDEs~\cite{Paeckel2019}, and they have even permeated to other fields, such as kinetic plasma simulation~\cite{Ye2022}. Time evolution problems usually rely on Fourier techniques with Trotter expansions~\cite{Greene2017, Lyu2022} or Chebyshev propagation schemes~\cite{Soley2017}. As an alternative, this work extends the HDAF approach to MPS/QTT, accurately and efficiently encoding differential operators, such as arbitrary functions of derivatives. Applied to the free-propagator, it enables Trotter expansions without Fourier transforms.

The work is structured as follows.
Section~\ref{sec:problem} presents the expansion problems to validate the numerical methods.
Section~\ref{sec:representation} introduces the MPS finite-precision algebra framework for quantum-inspired numerical algorithms and the HDAF machinery to approximate derivatives and functions of derivatives as MPOs, including metaheuristics to tune these approximations.
Section~\ref{sec:TimeEvolution} reviews the quantum time evolution problem and presents the numerical methods to address it. These techniques are contrasted via a one-step study on the one-dimensional quantum quench problem to determine the most convenient option.
Then, Section~\ref{sec:application} examines the best-performing technique for a long-time evolution on the complete range of expansion for the harmonic and non-harmonic problems.
Finally, Section~\ref{sec:conclusions} summarizes the conclusions of this study.

\section{Particle expansion}\label{sec:problem}

A quantum quench is a fundamental process where a system is driven out of equilibrium by a sudden change in its Hamiltonian~\cite{Mitra2018, Das2020}. In this section, we introduce the problem of particle expansion due to the sudden relaxation of an initial harmonic potential.

The particle expansion process presents several characteristics relevant to our study. First, it is a problem of interest in many areas, including many-body physics~\cite{Calabrese2006, Iucci2009, Calabrese2011, Essler2012} and quantum optomechanics~\cite{RodaLlordes2024}. Second, it stresses traditional and MPS-based numerical methods. Vector representations become too expensive for large expansions as they require the storage of a huge spatial domain. This makes the problem an interesting playground for MPS simulation. Nonetheless, the particle acceleration induces a chirping of the wavefunction, thus incrementing the bond dimension of the solution and putting a strain on MPS-based simulations as well. A third feature of this process is that when the quench varies from one harmonic potential to another, a closed-form analytical solution is known~\cite{GarciaRipoll_1999}. This is a critical feature allowing us to benchmark not only the speed but also the accuracy of our methods.

The Schrödinger equation describes the evolution of a particle's wavefunction
\begin{equation}\label{eq:Schrödinger}
  i\partial_t \psi(x,t) = \left(-\frac{\hbar^2}{2m}\partial_x^2 + V(x,t)\right)\psi(x,t).
\end{equation}

The election of the potential $V(x,t)$ determines the physical behavior of the system. The benchmark problem we introduce is the sudden change of a harmonic potential from frequency $\omega_{0}$ to $\omega_{H}$ at time $t=0$,


\begin{equation}\label{eq:ho_potential}
  V(x,t) = 
  \begin{cases}
    \frac{1}{2}\omega_0^2x^2,& t \leq 0,\\
    \frac{1}{2}\omega_H^2x^2,& t> 0.
  \end{cases}
\end{equation}

Assuming the wavefunction at time $t = 0$ is relaxed to the ground state of the previous Hamiltonian,
\begin{align}
  \label{eq:initial_wavefunction}
  \psi(x, t=0) &= \left(\frac{\omega_{0}}{\pi}\right)^{1/4}\exp\left(-\frac{1}{2}\omega_{0}x^{2}\right),
\end{align}
its evolution is prescribed according to
\begin{align}
  \psi(x,t) &= \left(\frac{\omega(t)}{\pi}\right)^{1/4}\exp\left(-\left[\frac{\omega(t)}{2} + i\beta(t)\right] x^{2}\right), \label{eq:analytic_solution} \\
  \omega(t) &= \omega_{H}\left(\frac{\omega_{H}}{\omega_0} \cos^2(\omega_H t)+\frac{\omega_0}{\omega_H} \sin^2(\omega_H t)\right)^{-1},\\
  \beta(t) &= \frac{\omega(t)}{4}\left(\frac{\omega_H}{\omega_0}-\frac{\omega_0}{\omega_H}\right)\sin(2\omega_H t).
\end{align}
The solution is a complex Gaussian with width ${\sigma(t)=1/\sqrt{\omega(t)}}$, modulated in time with period ${\pi/\omega_H}$. For ${\omega_H < \omega_0}$, the system undergoes an expansion during the first half of the period. The smallest width is ${\sigma_{\text{min}} = 1 / \sqrt{\omega_{0}}}$ at time ${t=0}$, and the largest width ${\sigma_{\text{max}} = \sqrt{\omega_{0}} / \omega_{H}}$ at time ${t = 0.5\pi / \omega_{H}}$.

The expansion is quantified by $\sigma_{\text{max}} / \sigma_{\text{min}} = \omega_0 / \omega_H$. That is, the frequency ratio $\omega_{0}/\omega_{H}$ is also the expansion ratio, dictating the total amplification of the wavefunction's spatial extent.
A larger expansion ratio requires more points to represent the solution accurately, challenging common computational approaches.

Another problem featuring similar expansion dynamics occurs when the initially harmonic potential changes to a wider trap with a non-harmonic component. This scenario holds high interest for experimental settings in optomechanics research~\cite{Levitodynamics, Setter2019, Neumeier2024, RodaLlordes2024b}. However,
analytical solutions are not usually known, while the intricacies that complicate numerical methods still hold.
One such case happens for a double-well potential,
\begin{equation}\label{eq:double_well}
  V(x,t) =
  \begin{cases}
    \frac{1}{2}\omega_0^2x^2, & t \leq 0,\\
    \frac{1}{2}\omega_H^2 x^2 + u \exp(-x^2/2\sigma^2), & t> 0.
  \end{cases}
\end{equation}

Qualitatively, the harmonic expansion behavior dominates the wavefunction's evolution for a sufficiently small value of $u$, with the Gaussian bump at ${x=0}$ acting as a perturbation. This allows us to use the results of the harmonic quench analysis as a basis for the study of the double-well potential. The Gaussian component of the potential, with ${u > 0}$, is expected to induce a symmetric separation of the expanding particle. This will deviate the wavefunction's evolution from the Gaussian form~\eqref{eq:analytic_solution} into a state with a two-peaked probability density. We choose this problem as a test to account for the feasibility of using the proposed numerical methods against a setting of actual research interest.

\section{MPS encoding and differential operators}
\label{sec:representation}
MPS originally arose in the domain of physics to study quantum many-body problems. Dolgov rediscovered this formalism in the field of applied mathematics under the name of quantized tensor trains (QTTs)~\cite{Dolgov2012}, a subclass of the broader class of tensor trains (TTs)~\cite{Oseledets2011}. A continuous alternative class of TTs is the functional tensor trains (FTTs)~\cite{Gorodetsky2019}, which hold a network of univariate matrix-valued functions instead of rank-three tensors. 

The exponential memory compression of MPS/QTT can bypass the curse of dimensionality in the representation of functions. This motivated the development of similar encodings from a quantum-inspired perspective~\cite{Lubasch2018, GarciaRipoll2021}. Indeed, MPS/QTT constitute efficient \textit{ansätze} for representing functions with fastly decaying Fourier coefficients~\cite{Jobst2023}.

MPS and TTs have been successfully applied to a variety of numerical analysis problems such as high-dimensional nonlinear PDEs~\cite{Dektor2021}, the Hamilton Jacobi Bellman equations~\cite{Horowitz2014, Stefansson2016, Gorodetsky2018, Dolgov2019, Oster2019}, the Schrödinger equation~\cite{Hong2022}, and stochastic problems~\cite{Lorenz2021, Dolgov2015, Eigel2016}. Combining Fourier techniques with Trotter expansions~\cite{Greene2017, Lyu2022}, or Chebyshev propagation schemes~\cite{Soley2017} allows for quantum dynamics simulations. Other approaches rely on one-step implicit time integration using an ALS-type solver or global space-time formulation to solve multi-dimensional parabolic problems~\cite{Dolgov2012}. Quantum-inspired proposals are successful at solving different PDEs, such as Schrödinger equations~\cite{Lubasch2018}, turbulence problems~\cite{Gourianov2022, gourianov_thesis}, Hamiltonian PDEs~\cite{GarciaMolina2022}, and the Vlasov-Poisson system~\cite{Ye2022}.

Combining this MPS-based representation of functions and the analogous encoding of operators as matrix product operators (MPO), along with a basic set of operations and efficient truncation algorithms, leads to a finite precision algebra~\cite{GarciaMolina2023} that operates similarly to standard matrix-vector operations. This algebra is the basis for developing time evolution quantum-inspired algorithms in section~\ref{sec:TimeEvolution}.

This section presents an extension of the Hermite Distributed Approximating Functionals (HDAF) to reconstruct functions of differential operators within this finite precision MPS-MPO framework. The HDAF performs this reconstruction as a linear combination of Hermite polynomials weighted by a Gaussian filter, approximating these operators with tunable pseudospectral precision at a limited cost. The section also covers a general review of the HDAF and the metaheuristics behind the use of this technique.

\subsection{Quantum-inspired numerical analysis}

The amplitude encoding of functions~\cite{GarciaRipoll2021,GarciaMolina2022,GonzalezConde2024} represents a function $f(x)$ with $x\in [a,b)$ in an $n$-qubit quantum register as a normalized quantum state
\begin{equation}\label{eq:function}
  \ket{f^{(n)}} = \frac{1}{\mathcal{N}_f^{1/2}} \sum_{s=0}^{2^n-1} f\left(x_s^{(n)}\right)\ket{s}, \quad x_s^{(n)} = a + s\Delta x ^{(n)},
\end{equation}
where $\mathcal{N}_f$ is the normalization constant and the index ${s=\lbrace 0,\dots, 2^n\rbrace}$ labels each coordinate $x_s$ to its corresponding quantum state $|s\rangle$~\cite{GarciaRipoll2021,GarciaMolina2022}. The amplitude encoding may be exponentially compressed in a quantum register by mapping the indices $s$ to the states of $n$ qubits. This leads to a binary encoding of the coordinates $s=\sum_{i=1}^{n}2^{n-i}s_i$, with $s_i = \lbrace 0, 1 \rbrace$.

This function representation in a quantum register creates a many-body wavefunction that, for bandwidth-limited functions, admits an efficient MPS representation~\cite{GarciaRipoll2021}
\begin{align}\label{eq:MPSrepresentation}
  \ket{\psi}&=\sum_{\lbrace s_i \rbrace}\sum_{\lbrace \alpha_i \rbrace} (A_{\alpha_1}^{s_1}A_{\alpha_{1},\alpha_2}^{s_2}\dots A_{\alpha_{N-1}}^{s_{N}})\ket{s_1}\\
            & \qquad\qquad\quad \otimes \ket{s_2} \otimes ... \otimes \ket{s_{N}},\notag
\end{align}
where each qubit is mapped to a site of the MPS. The physical indices $s_i$ correspond to the values of the energy levels of the qubits, and $\alpha_i$ are the bond indices, where $\alpha_i = \lbrace1,\dots,\chi_i\rbrace$ with $\chi_i$ the bond dimension.

Given this efficient representation of functions, a relevant challenge is to find a comparable representation for operators that act on these functions, from PDE operators that describe the action of derivatives and potentials to evolution operators that model the dynamics of the quantum state.
Potentials are diagonal operators and can be derived from the MPS representation of their corresponding functions either exactly or via interpolation techniques like the Chebyshev approximation~\cite{RodríguezAldavero2024} or the TT-cross interpolation~\cite{Oseledets2010, Dolgov2020}. Thus, the real challenge arises from
finding suitable, efficient representations of derivatives and functions thereof. 

The finite difference method is one of the most widespread techniques for approximating derivatives. This method uses the order $p$ Taylor expansion of the function, with an error $O(\Delta x^m)$, where $m$ depends on the terms combined for the approximation. Thus, the grid size limits the accuracy of the finite difference method. The most common implementations are the centered finite difference formulas
\begin{align} \label{eq:finite_difference_1}
  \frac{\partial f(x)}{\partial x} &= \frac{f(x + \Delta x) - f(x - \Delta x)}{2 \Delta x} + O(\Delta x^2), \\\label{eq:finite_difference_2}
  \frac{\partial^2 f(x)}{\partial x^2} &= \frac{f(x + \Delta x) - 2 f(x) + f(x - \Delta x)}{ \Delta x^2} + O(\Delta x^2),
\end{align}
whose truncation error scales quadratically with the discretization. The weak noise suppression of the centered finite difference formula can be enhanced to construct smooth noise-robust differentiators~\cite{Holoborodko2008}.

The finite difference MPO uses a linear combination of displacement operators $\Sigma^{\pm}$\cite{GarciaRipoll2021},
\begin{align}
  \ket{\partial_{x}f^{(n)}} &\simeq \frac{1}{2\Delta{x}}\left(\hat{\Sigma}^+-\hat{\Sigma}^-\right)\ket{f^{(n)}} + O(\Delta x^2), \\
  \ket{\partial^2_{x}f^{(n)}} &\simeq \frac{1}{\Delta{x}^2}\left(\hat{\Sigma}^+-2\mathbb{I}+\hat{\Sigma}^-\right)\ket{f^{(n)}} + O(\Delta x^2).
\end{align}
This MPO representation of the finite difference operators has a fixed bond dimension $\chi=3$ independent of the number of sites.

In addition to the truncation error, the round-off error affects the finite-difference scheme. The trade-off of truncation and round-off error determines the optimum step size $\Delta x$ since the truncation error decreases with $\Delta x$, but still, values that are too small will result in numerical errors due to round-off. Let us consider the second-order derivative approximation since this appears in the propagator of the time evolution. 
The round-off error occurs when the difference between two consecutive points $|f(x)-f(x\pm \Delta x)|$ of the discretized function is of the order of the machine precision $\delta$. This error is proportional to $\delta/\Delta x^2$ and can be amplified by the small denominator $\Delta x^2$. Increasing the step size while maintaining the number of points for the discretization can correct this error. 


\subsection{Hermite Distributed Approximating Functionals}

This work aims to overcome finite difference formulas' precision and speed limitations. In this regard, it is well known that spectral methods, such as Fourier techniques, can provide exponential speedup and precision guarantees for sufficiently smooth functions. Such methods have been derived in the domain of MPS/QTT methods~\cite{GarciaRipoll2021, Jobst2023, Chen2023, Dolgov_Khoromskij_Savostyanov_2012} with some ad-hoc heuristics to improve the creation of operators. Hermite Distributed Approximating Functionals (HDAF) give a powerful yet lesser-known numerical analysis technique. In the following, we will show how these methods work and how they can be adapted to engineer MPOs of arbitrary differential operators with relatively low costs. Our work differs from previous studies where they have been applied in matrix form to each site of a tensor train~\cite{DeGregorio_Iyengar_2019}.

The Distributed Approximating Functionals (DAF) are well-tempered approximations to the Dirac delta distribution.
The first DAF, developed before the class name was coined, was the Hermite DAF~\cite{Hoffman_Nayar_Sharafeddin_Kouri_1991},
\begin{align}
  \label{eq:hdaf-definition}
  \delta_{M}(x; \sigma) = \frac{\exp\left(\frac{-x^{2}}{2\sigma^{2}}\right)}{\sqrt{2\pi}\sigma}\sum_{m=0}^{M/2}\left(-\frac{1}{4}\right)^{m}\frac{H_{2m}\left(\frac{x}{\sqrt{2}\sigma}\right)}{m!},
\end{align}
where $H_{n}(x)$ is the $n$-th Hermite polynomial. It has two free parameters: The even integer $M$ and the positive real $\sigma$ are the order of the highest polynomial and the width of the approximation to the delta distribution, respectively.

The kernel defined in Eq.~\eqref{eq:hdaf-definition} is a nascent delta function that operates as the identity for polynomial functions of degree $M$ or lower,
\begin{align}
  \label{eq:hdaf-identity}
  f(x) \approx \int dx'\, \delta_{M}(x - x'; \sigma) f(x'),
\end{align}
approaching the Dirac distribution in the limit $\sigma/M \to 0$. However, unlike the exact delta distribution, $\delta_{M}(x; \sigma)$ is generally a bandwidth-limited, infinitely smooth function amenable to quadrature methods and differentiation.

The method's well-tempered property arises from the absence of special points in the reconstruction. Exact reproduction at grid points is not required; therefore, it does not constitute an interpolation scheme.
Moreover, a fundamental property of Eq.~\eqref{eq:hdaf-identity} is that the approximation converges uniformly to $f(x)$~\cite{Chandler_Gibson_1999}, and the approximation error usually resembles the function $f(x)$ albeit several orders of magnitude smaller~\cite{Hoffman_Kouri_1995}. This implies that the error is smaller when the function approaches zero, which is relevant and desirable for our wavefunctions applications.

Equation~\eqref{eq:hdaf-identity} is customarily discretized on a uniform grid with spacing $\Delta x$, using midpoint integration to render it as a matrix-vector product, $f(x_{i}) = \sum_{j}K_{ij}f(x_{j})$, where $K$ is a symmetric Toeplitz matrix with components
\begin{align}
  \label{eq:hdaf-matrix}
  K_{ij} = \Delta x \delta_{M}(\Delta x |i - j|; \sigma).
\end{align}

Since $\delta_{M}(x; \sigma)$ has an exponentially decaying envelope, two essential properties arise: (i) the reconstruction matrix $K$ can be made highly sparse and concentrated around its main diagonal, with the number of diagonals controlled by $\sigma / \Delta x$, and (ii) a minimal number of diagonals (i.e. quadrature nodes) are required to accurately discretize the integral in Eq.~\eqref{eq:hdaf-identity}. The composite midpoint rule converges especially fast for periodic or peaked functions with vanishing derivatives at the integration limits~\cite{Goodwin_1949}. Moreover, in this context, the midpoint rule surpasses the accuracy of higher-order Newton-Cotes quadrature rules using the same number of grid points~\cite{Dyson_1999, Weideman_2002}.

The MPO corresponding to the $K$ matrix can be constructed as a weighted combination of the displacement operators $\Sigma^\pm$ as
\begin{align}
  \hat{K} &= \Delta x \delta_{M}(0; \sigma) \mathbb{I} \nonumber \\
          & \qquad +  \sum_{i=1}^{2^{n} - 1} \Delta x \delta_{M}(i\Delta x; \sigma) \left(\hat{\Sigma}^{+i} + \hat{\Sigma}^{-i}\right),
            \label{eq:hdaf-mpo}
\end{align}
where the symmetry of $\delta_{M}$ has been used. While in principle the sum ranges over the whole grid, in practice, one only needs to sum until $\delta_{M}$ has vanished according to a prescribed tolerance, as detailed on Sec~\ref{sec:metaheur-summ-bounds}.

\subsection{HDAF differentiation}
\label{sec:hdaf-differentiation}

The HDAF formalism opens a path to estimate a function's derivative of any order, as well as functions $D[\partial/\partial x]$ of such derivatives. From Eq.~\eqref{eq:hdaf-identity} it follows that
\begin{align}
  \label{eq:hdaf-reconstruction-differential-operator}
  D\left[\frac{\partial}{\partial x}\right] f(x) \approx \int dx'\, D\left[\frac{\partial}{\partial x}\right] \delta_{M}(x - x') f(x').
\end{align}
An analytical expression for $D[\partial/\partial x] \delta_{M}(x - x')$ is usually easy to find. The typical procedure involves using the Rodrigues formula for the Hermite polynomials,
\begin{align}
  \label{eq:hermite-rodrigues-formula}
  H_{n}(x) = (-1)^{n} \exp(x^{2})\frac{\partial^{n}}{\partial x^{n}}\exp(-x^{2}),
\end{align}
to rewrite Eq.~\eqref{eq:hdaf-definition} as
\begin{align}
  \label{eq:hdaf-for-calculations}
  \delta_{M}(x; \sigma) &= \frac{1}{\sqrt{2\pi}\sigma}\sum_{m=0}^{M/2}\frac{1}{m!} \left(-\frac{\sigma^{2}}{4}\right)^{m} \nonumber \\
                        & \qquad\qquad\qquad \times \frac{\partial^{2m}}{\partial x^{2m}}\exp\left(-\frac{x^{2}}{2\sigma^{2}}\right).
\end{align}
Then, the differential operator is applied over ${\delta_{M}(x - x'; \sigma)}$ acting on the exponential, and the Rodrigues formula~\eqref{eq:hermite-rodrigues-formula} is used to recover an expression in terms of Hermite polynomials, without explicit derivatives. For instance, the $l$-th derivative leads to
\begin{align}
  \label{eq:hdaf-derivatives}
  \delta_{M}^{(l)}(x; \sigma) &= \left(\frac{-1}{\sqrt{2}\sigma}\right)^{l} \frac{\exp\left(\frac{-x^{2}}{2\sigma^{2}}\right)}{\sqrt{2\pi}\sigma} \nonumber \\
                              & \qquad \times \sum_{m=0}^{M/2}\left(-\frac{1}{4}\right)^{m} \frac{H_{2m + l}\left(\frac{x}{\sqrt{2}\sigma}\right)}{m!}.
\end{align}

Analog to the reconstruction~\eqref{eq:hdaf-mpo}, the differentiating MPO for the $l$-th derivative in the HDAF formalism is
\begin{align}
  \label{eq:hdaf-derivatives-mpo}
  \hat{K}^{(l)} &= \Delta x \delta_{M}^{(l)}(0; \sigma) \mathbb{I} \nonumber \\
                & \quad +  \sum_{i=1}^{2^{n} - 1} \Delta x \delta_{M}^{(l)}(i\Delta x; \sigma) \left(\hat{\Sigma}^{+i} + (-1)^{l}\hat{\Sigma}^{-i}\right),
\end{align}
where the symmetry or antisymmetry of $\delta_{M}^{(l)}$ has been used for $l$ even or odd, respectively.

Attention must be paid to the fact that the $l-$th derivative has a prefactor $\sigma^{-(l+1)}$ with $\sigma = O(\Delta x)$. Since these operators will be used within a finite-precision numerical framework, round-off errors can significantly impact the accuracy of the differentiation. These errors arise from the limited significant digits that computers may represent, inducing small approximation errors amplified by large weighting prefactors. In particular, these deviations dominate when $\Delta x$ is too small, thus limiting how dense the numerical grid can be.

The round-off error amplification problem is expected in numerical differentiation techniques but can be mitigated.
In the particular MPS-MPO framework, differentiating operators can be discretized to accommodate a certain number of qubits and then be adapted to a denser grid, where identities are added as new sites to account for each additional qubit. This approach is equivalent to nearest-neighbor interpolation. It keeps the round-off errors constant and does not introduce extra complexity to the MPO.

\begin{figure}[ht!]
  \centering
  \includegraphics[width=1\linewidth]{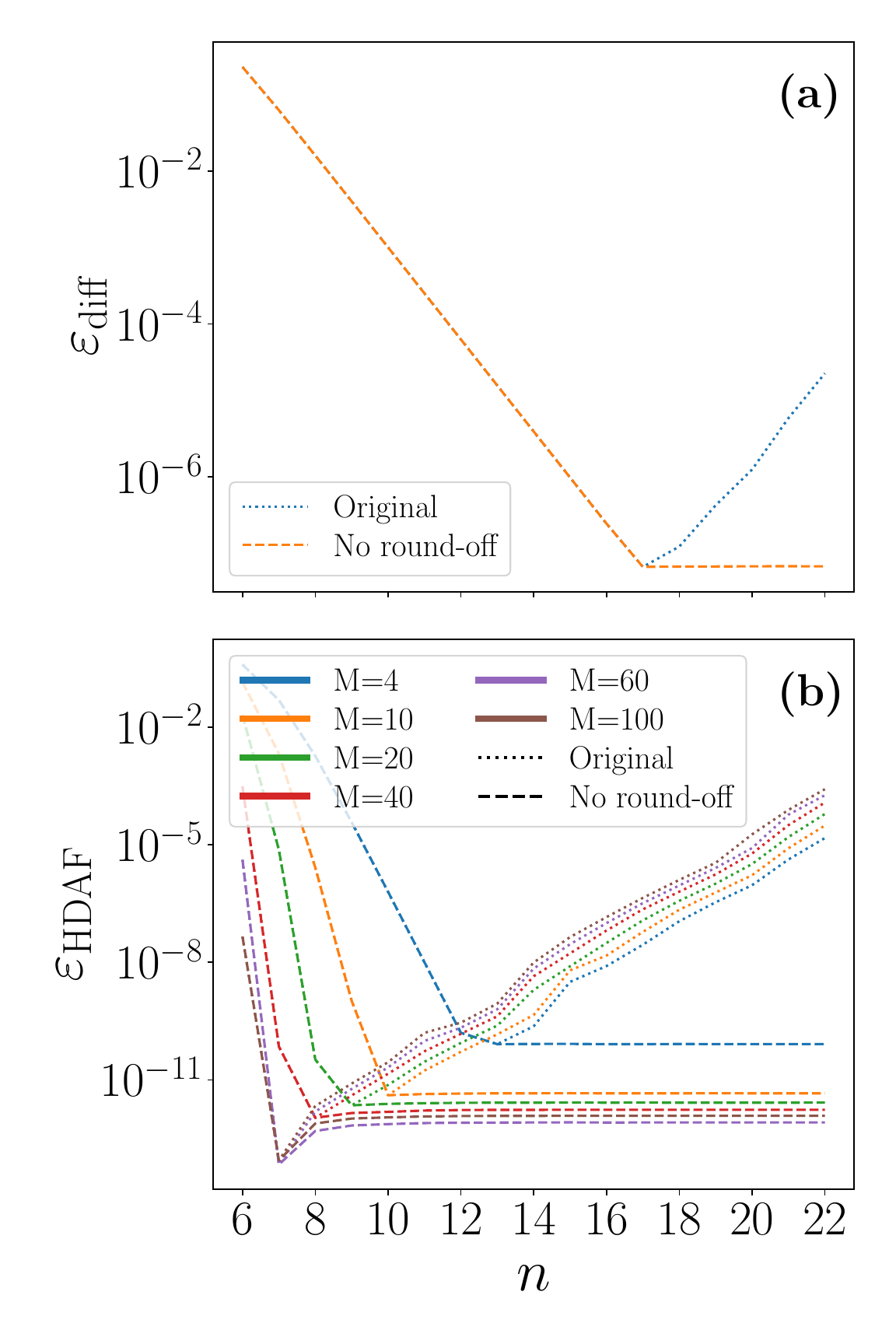}
  \caption{Errors in the second derivative approximation of a Gaussian function for a varying number of qubits. Dotted lines correspond to the direct implementation of the differentiating operators. Dashed lines implement the procedure specified at the end of Sec.~\ref{sec:hdaf-differentiation} to limit round-off errors.
    (a) Finite differences, (b) HDAF.}
  \label{fig:derivative_error}
\end{figure}

Figure~\ref{fig:derivative_error} accounts for the differentiation accuracy of HDAF and finite differences in approximating the second derivative of a Gaussian function. In the case of the HDAF, the convergence is faster in the number of qubits for larger values of $M$, as expected. However, HDAF and finite differences suffer from relevant round-off errors for many qubits. Once the optimal accuracy is achieved for a certain number of qubits, the accuracy of the operators can be retained while acting on a finer grid. Round-off errors are effectively kept constant by following the procedure above.

\subsection{HDAF free propagator}

The HDAF scheme is very powerful, as it cannot only approximate derivatives but also general functions of those derivatives. The first and one of the most relevant applications of this idea, posed in Ref.~\cite{Hoffman_Nayar_Sharafeddin_Kouri_1991}, is the
banded approximation of the free propagator.

From Eq.~\eqref{eq:hdaf-reconstruction-differential-operator}, taking the differential operator $D[\partial/\partial x]$ to be the free propagator,
\begin{align}
  \label{eq:free-propagator}
  T(\tau) = \exp\left(-\frac{i\tau}{2}\frac{\partial^{2}}{\partial x^{2}}\right),
\end{align}
the kernel for the approximation is $T(\tau)\delta_{M}(x - x'; \sigma)$.
This quantity is readily computed from Eq.~\eqref{eq:hdaf-for-calculations} since $T(\tau)$ commutes with the derivatives and it spreads a Gaussian function to another,
\begin{align}
  T(\tau)\exp\left(\frac{-(x - x')^{2}}{2\sigma^{2}}\right) &= \left[\frac{\sigma}{\sigma_{\tau}}\right]\exp\left(\frac{-(x-x')^{2}}{2\sigma_{\tau}^{2}}\right),
\end{align}
mapping the original variance $\sigma^{2}$ to ${\sigma_{\tau}^{2} = \sigma^{2} + i\tau}$. Then, using Eq.~\eqref{eq:hermite-rodrigues-formula} yields the free propagator kernel
\begin{align}
  \delta_{M}(x - x'; \sigma, \tau) &= T(\tau) \delta_{M}(x - x'; \sigma) \nonumber \\
                                   &= \frac{\exp\left(\frac{-(x-x')^{2}}{2\sigma_{\tau}^{2}}\right)}{\sqrt{2\pi}\sigma_{\tau}} \nonumber \\
                                   & \quad \times \sum_{m=0}^{M/2}\left(\frac{-\sigma^{2}}{4\sigma_{\tau}^{2}}\right)^{m}\frac{H_{2m}\left(\frac{x - x'}{\sqrt{2}\sigma_{\tau}}\right)}{m!}.
                                     \label{eq:hdaf-free-propagator}
\end{align}

The MPO for the free propagator in the HDAF formalism is
\begin{align}
  \label{eq:hdaf-free-propagator-mpo}
  \hat{K}_{\tau} &= \Delta x \delta_{M}(0; \sigma, \tau) \mathbb{I} \nonumber \\
                 & \qquad +  \sum_{i=1}^{2^{n} - 1} \Delta x \delta_{M}(i\Delta x; \sigma, \tau) \left(\hat{\Sigma}^{+i} + \hat{\Sigma}^{-i}\right).
\end{align}

The kernel~\eqref{eq:hdaf-free-propagator} becomes complex and highly oscillating as time increases. Also, its width increases as a fundamental consequence of the free propagator. However, the spreading in the HDAF formalism is the minimum possible since it is inherited from the Gaussian generator of the Hermite polynomials~\cite{Hoffman_Kouri_1992}.

The HDAF approximation for the propagator has been used in many applications of split-step integration methods within the traditional vector framework. In that framework, $\hat{K}_{\tau}$ is represented as a matrix acting on a discretized function. A central contribution in this work is to realize that the same matrix can be more efficiently represented as an MPO using the displacement operators $\hat{\Sigma}^{\pm}$ and additional simplification steps that significantly reduce the effective bond dimension of the operator. In this scenario, the MPO HDAF propagator is a competitive alternative to using MPO Fourier-based techniques~\cite{GarciaRipoll2021}, directly representing the evolution operator in the coordinate representation.

\subsection{HDAF metaheuristics}

\subsubsection{Free parameter election}
\label{sec:free-parameter-election}

\paragraph{Identical reconstruction.}

The formulation of the HDAF operator~\eqref{eq:hdaf-identity} as a discrete matrix has $2$ sources of error: (i) The assumption that the function $f(x)$ can be expressed as a polynomial of degree $M$ under the extent of the Gaussian envelope of the HDAF, and (ii) the discretization of the convolution integral to a finite sum employing the midpoint rule. While the error (i) vanishes in the limit $M/\sigma \to \infty$, it is clear from (ii) that it is not possible to increase $M$ or decrease $\sigma$ indefinitely. A more oscillatory integrand will require a larger value of $\sigma / \Delta x$ for the Gaussian envelope to cover enough nodes and achieve satisfactory integration accuracy.

In general, for a fixed $M$, there is a value of $\sigma/\Delta x$ that makes the reconstruction optimal, and the larger is $M$, the better the maximum accuracy that can be achieved. The rationale behind this optimal relationship between $M$ and $\sigma/\Delta x$ is that a perfect reconstruction happens when the $M$ zeroes of the HDAF match with the zeroes on the grid, and only the term in the origin contributes~\cite{Hoffman_Arnold_Kouri_1992}. One possible approach, therefore, is to set the discrete HDAF to be $1$ at the origin~\cite{Pindza_Maré_2014},
\begin{align*}
  K_{ii} = \Delta x \delta_{M}(0; \sigma) = 1,
\end{align*}
yielding,
\begin{align}
  \label{eq:sigma-from-M}
  \sigma_M = \frac{\Delta x}{\sqrt{2\pi}}\sum_{m=0}^{M/2}\left(\frac{-1}{4}\right)^{m}\frac{H_{2m}(0)}{m!},
\end{align}
which makes the HDAF approximately vanish at integer multiples of $\Delta x$~\cite{Mazzoni_2014}.

In practice, a lower bound to $\sigma/\Delta x$ is prescribed to ensure convergence of the midpoint rule when $M$ is small. For the context of double floating-point precision, we heuristically set
\begin{align}
  \label{eq:hdaf-minimum-sigma}
  \sigma / \Delta x \geq 3
  \Rightarrow
  \sigma_{\text{min}} = 3\Delta x,
\end{align}
and choose the value of $\sigma$ according to
\begin{align}
  \label{eq:hdaf-identity-calibration}
  \sigma = \max( \sigma_{M}, \sigma_{\text{min}}).
\end{align}

\paragraph{Differentiation.}

Assuming that the $l-$th derivative of the function $f(x)$ is accurately described with an HDAF of order $M$, i.e., also pertains to the DAF-class~\cite{Hoffman_Kouri_1995}, then the approximation to the derivative can be thought as a reconstruction of $f^{(l)}(x)$ instead of $f(x)$,
\begin{align*}
  f^{(l)}(x) &\approx \int dx'\, \delta_{M}^{(l)}(x - x'; \sigma) f(x') \\
             &= \int dx'\, \delta_{M}(x - x'; \sigma) f^{(l)}(x').
\end{align*}

In this spirit, the optimal value of $\sigma$ is again computed using Eq.~\eqref{eq:hdaf-identity-calibration}, which does not depend upon the function to reconstruct, provided $M$ is fixed.

\paragraph{Free evolution.}

Since the action of the free propagator is to spread the original wavefunction, the width of the HDAF will not be a problem for the midpoint integration. For efficiency purposes, the election of $\sigma$ is made such that the new width of the freely-propagated HDAF is the smallest possible~\cite{Hoffman_Nayar_Sharafeddin_Kouri_1991}. This value follows from equation~\eqref{eq:hdaf-free-propagator}. The leading Gaussian,
\begin{align*}
  \exp\left(\frac{-x^{2}}{2(\sigma^{2} + i\tau)}\right) =
  \exp\left(\frac{-x^{2}}{2 w^{2}}\right)
  \exp\left(i\frac{x^{2}}{2w^{2}}\frac{\tau}{\sigma^{2}}\right),
\end{align*}
has an effective variance $w^{2} = (\sigma^{2} + \tau^{2}/\sigma^{2})$ with an optimal value $\sigma = \sqrt{\tau}$ that minimizes its spatial extent. Then, the value of $\sigma$ is chosen
\begin{align}
  \label{eq:hdaf-propagator-calibration}
  \sigma = \max( \sigma_{M}, \sigma_{\text{min}}, \sqrt{\tau}),
\end{align}
where $\sigma_{M}$ and $\sigma_{\text{min}}$ and are the same values~\eqref{eq:sigma-from-M} and~\eqref{eq:hdaf-minimum-sigma} used for identical HDAF reconstruction.

\subsubsection{Self-consistent error estimation}
\label{sec:self-cons-error}

The HDAF filter~\eqref{eq:hdaf-identity} can be analyzed in Fourier space. From equation~\eqref{eq:hdaf-for-calculations}, the kernel spectrum has the analytical form
\begin{align}
  \label{eq:hdaf-fourier-transform}
  \widehat{\delta}_{M}(k; \sigma) = \exp\left(\frac{-k^{2}\sigma^{2}}{2}\right)\sum_{m=0}^{M/2}\frac{1}{m!}\left(\frac{k^{2}\sigma^{2}}{2}\right)^{m}.
\end{align}
The summation is a truncated series expansion of $\exp(k^{2}\sigma^{2}/2)$ up to order $M/2$. This expression is the basis to prove that the HDAF filter approaches a true Dirac delta distribution in the limits of an infinitely broad filter or an infinitely large polynomial basis,
\begin{align*}
  \lim_{\sigma\to 0} \widehat{\delta}_{M}(k; \sigma) = \lim_{M\to\infty} \widehat{\delta}_{M}(k; \sigma) = 1 \quad \forall k\in\mathbb{R},
\end{align*}
thus identically preserving the function to reconstruct.

From equation~\eqref{eq:hdaf-fourier-transform}, one can also note that the Fourier expression of the HDAF is symmetric, bounded, and monotonically decreasing in ${k\in\mathbb{R^{+}}}$, with
\begin{align*}
  1 = \widehat{\delta}_{M}(0; \sigma) \ge \widehat{\delta}_{M}(k; \sigma) \ge \lim_{k\to\infty}\widehat{\delta}_{M}(k; \sigma) = 0,
\end{align*}
therefore acting as a low-pass filter. Moreover, it has been shown that ${\widehat{\delta}_{M}(k, \sigma)}$ is an almost-ideal low-pass filter with transition frequency
\begin{align}
  \label{eq:hdaf-transition-frequency}
  k^{\star} = \sqrt{M + 1} / \sigma,
\end{align}
and a transition region width that scales as ${O(M^{-1/2}\sigma^{-1})}$~\cite{Bodmann_Hoffman_Kouri_Papadakis_2008}.

\begin{figure}[ht]
  \centering
  \includegraphics[width=\linewidth]{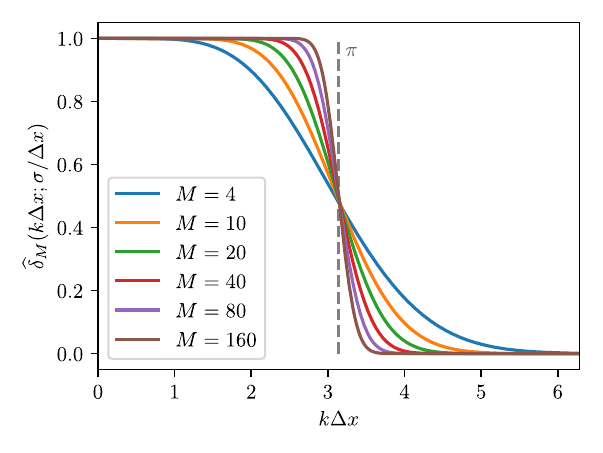}
  \caption{Fourier spectrum of $\delta_{M}(x; \sigma)$. Frequencies and widths are normalized to the grid spacing $\Delta x$. The width $\sigma$ is computed for each $M$ according to equation~\eqref{eq:sigma-from-M}.}
  \label{fig:hdaf-filter}
\end{figure}

This behavior is depicted in Figure~\ref{fig:hdaf-filter} for varying $M$. There is a region of frequencies below $k^{\star}$, the so-called DAF-plateau, such that ${\widehat{\delta}_{M}(k; \sigma) \approx 1}$. There follows a transition region centered around $k^{\star}$ where the value of the filter smoothly vanishes, and then it indefinitely approximates to zero. As $M$ is larger, the DAF plateau extends closer to $k^{\star}$. Note that choosing $\sigma$ from equation~\eqref{eq:sigma-from-M} fixes the transition frequency ${k^{\star} \approx \pi/\Delta x}$, which is the maximum frequency representable on a discrete uniform grid.

The HDAF reconstruction will be accurate for bandwidth-limited functions whose spectra lie within the DAF plateau, and any function with higher-frequency contributions will be smoothed. This suggests that this formalism is a good fit to use along with matrix product states since both techniques are especially suitable to represent bandwidth-limited functions~\cite{Jobst2023}.

\subsubsection{Evaluation of the HDAF coefficients.}

All the HDAF operators presented here are generated as a combination of displacements and coefficients
\begin{align}
  \label{eq:hdaf-general-operator}
  K^{(l)}_{\tau} = \sum_{k=-2^{n}+1}^{2^{n}-1} \Delta x \delta_{M}^{(l)}(k\Delta x; \sigma, \tau)\left(\Sigma^{+}\right)^{k}.
\end{align}
Moreover, the coefficients fulfill the general form
\begin{align}
  \label{eq:hdaf-general-coefficients}
  \Delta x\delta_{M}^{(l)}(x; \sigma, \tau)  &= d_{l}\sum_{m=0}^{M/2}h_{m, l}(x / \sqrt{2(\sigma^{2} + i\tau)}),
\end{align}
with the definitions
\begin{align*}
  d_{l} &= \frac{(-1)^{l} \Delta x}{\sqrt{2(\sigma^{2} + i\tau)}^{l+1}\sqrt{\pi}},\\
  h_{n, l}(x) &= H_{2n + l}(x)\exp(-x^{2})\frac{c^{n}}{n!}, \\
  c &= -\frac{1}{4}\frac{\sigma^{2}}{(\sigma^{2} + i\tau)}.
\end{align*}

From the properties of Hermite polynomials, it follows that $h_{n, l}(x)$ obeys the double recurrence relation
\begin{align*}
  h_{n+1, l}(x) &= \frac{2c}{n + 1}\left[ x h_{n, l+1}(x) - (2n + l + 1)h_{n, l}(x) \right],\\
  h_{n+1, l+1}(x) &= 2x h_{n+1, l}(x) - 2c(2 + l/(n+1))h_{n, l+1}(x),
\end{align*}
that makes the calculation of~\eqref{eq:hdaf-general-coefficients} efficient and accurate, starting from the initial values $h_{0, l}(x) = H_{l}(x)\exp(-x^{2})$ and $h_{0, l+1}(x) = H_{l+1}(x)\exp(-x^{2})$.

\subsubsection{Effective summation bounds.}
\label{sec:metaheur-summ-bounds}

It was previously mentioned that equations~\eqref{eq:hdaf-mpo},~\eqref{eq:hdaf-derivatives-mpo},~\eqref{eq:hdaf-free-propagator-mpo} and~\eqref{eq:hdaf-general-operator} formally sum over all the grid points of $x$. In practice, however, the fast decay of the HDAF narrows the sum to a small subset of points around the origin. This subset can be further restricted since the filters are either symmetric or antisymmetric.

Only the highest power on the argument of ${\delta_{M}^{(l)}(x; \sigma, \tau)}$ will contribute significantly to the value of the HDAF for ${x \gg 0}$.
Let $W$ be the smallest positive integer such that the coefficients contribute at most a predefined error tolerance $\varepsilon$,
\begin{align*}
  \left|\Delta x \delta_{M}^{(l)}(W\Delta x; \sigma, \tau)\right| \leq \varepsilon.
\end{align*}

This integer can be estimated tightly by replacing the sum in~\eqref{eq:hdaf-general-coefficients} with the highest power term on ${W\Delta x}$ from the polynomial on ${h_{M/2, l}(W\Delta x / \sqrt{2(\sigma^{2} + i\tau)})}$, but it leads to a transcendental equation for $W$. Instead, we approximate the sum by its last term as a whole and use the following upper bound for the Hermite polynomials with complex argument~\cite{vanEijndhoven1990},
\begin{align*}
  \left| H_{n}(z) \right| \leq \sqrt{2}^{n}\sqrt{n!}\exp(\sqrt{2n}|z|), \quad z\in\mathbb{C}, n\in\mathbb{N}.
\end{align*}

While this bound tends to overestimate the polynomial, the Gaussian envolvent will dominate fast enough for this not to become a significant problem. Then, we find $W$ by setting
\begin{align*}
  |\Delta x\delta_{M}^{(l)}| &\lesssim \frac{\Delta x\sqrt{(M+l)!}}{\sqrt{2\pi}\sqrt{|\sigma^{2} + i\tau|}^{l+1}(M/2)!} \left|\frac{\sigma^{2}}{2 (\sigma^{2} + i\tau)}\right|^{M/2} \nonumber \\
                             &\quad\times \exp\left( W\Delta x\sqrt{\frac{M + l}{|\sigma^{2} + i\tau|}}\right) \nonumber \\
                             &\quad \times \exp\left(- \frac{W^{2}\Delta x^{2}}{2(\sigma^{2} + \tau^{2}/\sigma^{2})} \right) \nonumber \\
                             &= \varepsilon,
\end{align*}
which reduces to the quadratic equation,
\begin{align}
  \label{eq:width-bound-equation}
  W^{2} \frac{\Delta x^{2}}{2(\sigma^{2} + \tau^{2}/\sigma^{2})} - W\Delta x\sqrt{\frac{M + l}{|\sigma^{2} + i\tau|}} + \ln\frac{\varepsilon}{\eta} = 0,
\end{align}
with

\begin{align*}
  \eta = \frac{\Delta x\sqrt{(M+l)!}}{\sqrt{2\pi}\sqrt{|\sigma^{2} + i\tau|}^{l+1}(M/2)!} \left|\frac{\sigma^{2}}{2 (\sigma^{2} + i\tau)}\right|^{M/2}.
\end{align*}

The HDAF MPOs~\eqref{eq:hdaf-general-operator} are obtained by summing over indices $-W$ to $W$, where $W$ is the closest integer from above to the solution of~\eqref{eq:width-bound-equation}. This sum contains $2W + 1$ weighted displacement operators $\hat{\Sigma}^{\pm k}$, where only $W + 1$ coefficients must be explicitly computed due to the symmetry $\delta_{M}^{(l)}(-x) = (-1)^{l}\delta_{M}^{(l)}(x)$. Despite the number of summands, the resulting MPO will, in practice, be relatively simple, with a small bond dimension, for a reasonable choice of the HDAF parameters.

\section{Time evolution algorithms}\label{sec:TimeEvolution}

Our goal is to solve the time-dependent Schrödinger equation~\eqref{eq:Schrödinger} with ${H=D(-\partial_x^2)+V(x)}$. The formal solution for this problem can be expressed as the repeated action of a possibly time-dependent unitary operator $U(t)$ on an initial state $\psi(x,t=0)$,
\begin{equation}\label{eq:evolution}
  \psi(x,t)=U(t)\psi(x,0)=e^{-i\hat{H}t}\psi(x,0).
\end{equation}

In the particular framework of problems we are interested in, the state $\psi(x,t)$ will be encoded using MPS/QTT, and we will use MPOs and finite-precision MPS algebra tools to approximate the unitary operator $U(t)$ for brief time periods. More explicitly, the studies below will either use an MPO structure to encode the Hamiltonian $H$ or create an MPO that directly approximates $U(t)$. In both cases, the representations based on QTT/MPS will provide us access to exponentially dense grids with $2^n$ points in space, which may prove advantageous regarding vector representations.

To address the time evolution problem, the PDE operators described in Section~\ref{sec:representation}
require global evolution schemes independent of the locality of interactions. MPS algorithms present many suitable alternatives, such as the time-dependent variational principle (TDVP)~\cite{Haegeman2011, Vanderstraeten2019, Paeckel2019}, and
Taylor, Padé, and Arnoldi approximations of the evolution operator~\cite{GarciaRipoll2006}. This section presents a selection of time-evolution methods with an MPO-MPS implementation: explicit (Euler, Improved Euler, and fourth-order Runge-Kutta) and implicit (Crank-Nicolson) Runge-Kutta methods, restarted Arnoldi iteration, and the split-step method. HDAF's propagator approximation enables the use of the split-step method. The rest of the methods are suitable for a finite difference and HDAF approximation of the differential operator.

\subsection{Runge-Kutta methods}\label{sec:RK}
Runge-Kutta methods approximate the time evolution by a Taylor expansion of the state with a local error, i.e., one-step error, that scales algebraically with the expansion order $m$ as $O(\Delta t^{m+1})$. Let us describe some of the most representative variations.

1. \underline{Euler method.}  The simplest order one method
\begin{align}
  \begin{split}
    \psi_0 &= \psi(x,t_0),  \\
    \psi_{k+1} &= \psi_k - i\Delta t H \psi_k, \quad \mbox{for } k=0,1,\dots,N-1.
  \end{split}
\end{align}

2. \underline{Improved Euler or Heun method.} This method improves on the Euler method with a second-order error given by
\begin{equation}
  \psi_{k+1} = \psi_k - i \frac{\Delta t}{2} \left[v_1 + H (\psi_k-i\Delta t v_1)\right],
\end{equation}
with $v_1 = H \psi_k$.

3. \underline{Fourth-order Runge-Kutta method.} Finally, the well-known fourth-order scheme
\begin{align}
  \psi_{k+1} &=\psi_k + i\frac{\Delta t}{6}(v_1+2v_2+2v_3+v_4),\;\mbox{with} \\
  v_1 &= - H \psi_k, \nonumber \\
  v_2 &= - H\left(\psi_k+i\frac{\Delta t}{2}v_1\right), \nonumber \\
  v_3 &= - H\left(\psi_k+i\frac{\Delta t}{2}v_2\right), \nonumber \\
  v_4 &= - H\left(\psi_k+i\Delta t v_3\right).\nonumber
\end{align}
This is one of the most commonly used methods in solving PDEs due to its balance in accuracy, stability, and simplicity.

4. \underline{Crank-Nicolson method.} Implicit methods can increase numerical stability. The Crank-Nicolson algorithm is a second-order implicit method based on the trapezoidal rule that combines the Euler method and its backward version evaluated on the $k$ and $k+1$ iterations, respectively. Thus, the state at the $k+1$ iteration is approximated as
\begin{equation}
  \left(\mathbb{I}+\frac{i\Delta t}{2}H\right)\psi_{k+1}=\left(\mathbb{I}-\frac{i\Delta t}{2}H\right)\psi_{k}.
\end{equation}
Matrix inversion methods may solve the system of equations in its matrix-vector implementation. Other approaches, such as conjugate gradient descent, can be extended for its implementation in an MPO-MPS framework.

\subsection{Restarted Arnoldi iteration}\label{sec:Arnoldi}
Another alternative is to use Krylov subspace methods; more concretely, the restarted Arnoldi iteration adapted to the time evolution problem. These methods rely on the Krylov basis $\mathcal{K}_L = \mathrm{lin}\{\ket{\psi_k}, H\ket{\psi_{k}},\ldots,H^{L-1}\ket{\psi_{k}}\}$ to construct an approximation of the evolution. 

The restarted Arnoldi iteration constructs a Krylov basis $\lbrace v_i\rbrace_{i=1,\dots,n_v}$ of $n_v$ elements and computes the matrices of the expectation value of the operator $H$ and its norm---$A$ and $N$, whose matrix elements are $\langle v_i|H| v_j\rangle$ and $\langle v_i| v_j\rangle$, respectively--- to approximate the exact exponential evolution as
\begin{equation}\label{eq:Arnoldi}
  \psi_{k+1}=e^{-i\Delta t N^{-1}A}\psi_k.
\end{equation}
The error in approximating the exponential function is limited by the number of Krylov vectors, which scales with $O(\Delta t^{n_v})$. This means that even a low number of vectors---$n_v=5,10$---can provide a highly accurate approximation. As a result, the cost of matrix inversion and exponentiation decreases significantly compared to the exact application of the $H$ operator since the dimensions of $A$ and $N$ can remain constant with the system size, thus avoiding exponential scaling.

\subsection{Split-step method}\label{sec:split-step}
Split-step methods are based on an approximate decomposition of the Hamiltonian exponential as a product of exponentials that can be efficiently computed.

The first-order method relies on the Lie-Trotter product formula to approximate the evolution operator as
\begin{equation}
  U(\Delta t) = e^{-i \Delta t D(-\partial_x^2)} e^{-i\Delta t V(x)} + O(\Delta t^2).
\end{equation}
Higher order expansions, such as the Suzuki-Trotter formulas~\cite{Suzuki1990, Suzuki1991}, enable a better approximation by decreasing the error scaling with the time step. A common alternative is the second-order approximation
\begin{align}
\begin{split}\label{eq:split-step}
    e^{-i \Delta t (D(-\partial_x^2)+V(x))} 
    &= e^{-i\Delta t V(x)/2} e^{-i \Delta t D(-\partial_x^2)}  \\
    & \quad \times e^{-i \Delta t V(x)/2} + O(\Delta t^3).
\end{split}
\end{align}

This decomposition generates the Störmer-Verlet integration scheme, the most common and one of the simplest members of the class of symplectic integrators~\cite{Yoshida_1992}. These integrators are designed to preserve the system's energy and are especially well-fitted for conservative long-time evolution problems. While most implementations rely on second-order accuracy in $\Delta t$, higher-order schemes can be constructed at the expense of introducing a larger number of exponential operators~\cite{McLachlan1992}.

Usually, the split-step~\eqref{eq:split-step} relies on two Fourier transforms to compute the action of the free propagator ${\exp(-i\Delta t D(-\partial_{x}^{2}))}$, but the HDAF formalism permits the efficient alternative of applying the free-propagator approximation~\eqref{eq:hdaf-free-propagator} directly in the coordinate representation.

The potential propagator $\exp(-i\frac{\Delta t}{2}V(x))$, diagonal in the coordinate basis, is approximated using TT-cross interpolation as implemented in Ref.~\cite{RodríguezAldavero2024}.
While the propagator for the harmonic potential can be cast analytically as an MPO, other potentials such as Eq.~\eqref{eq:double_well} do not have an exact representation and must be computed numerically.

\subsection{One-step study}\label{sec:one-step}

\begin{figure*}[t!]
  \centering
  \includegraphics[width=0.9\linewidth]{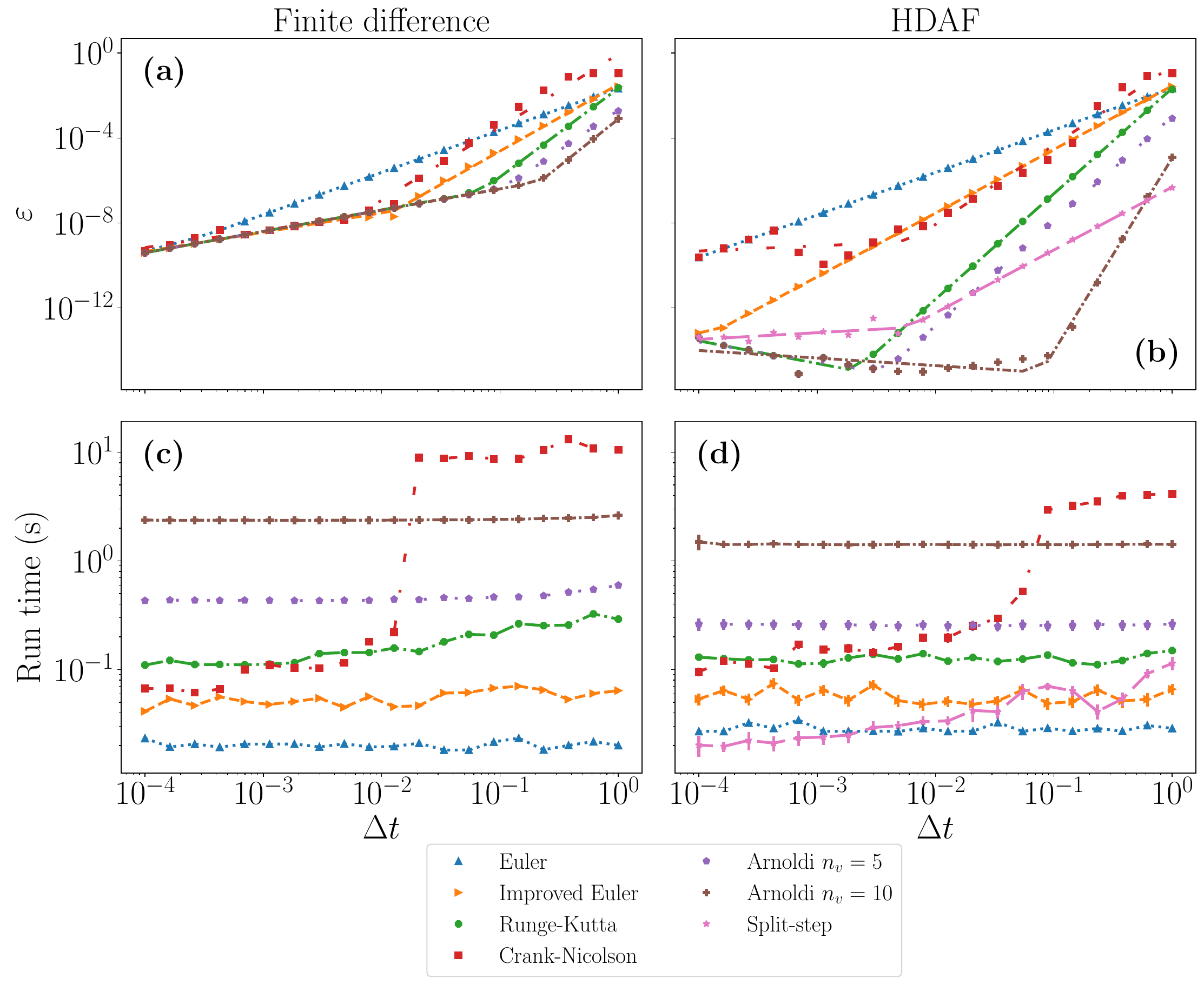}
  \caption{One-step evolution for a range of $\Delta t$ and a fixed number of qubits $n=18$. (a) Error $\varepsilon$ (finite difference), (b) Error $\varepsilon$ (HDAF), (c) run time (finite difference), (d) run time (HDAF). The run time is averaged over ten runs.}
  \label{fig:FD_vs_HDAF}
\end{figure*}

\begin{figure}
  \centering
  \includegraphics[width=\linewidth]{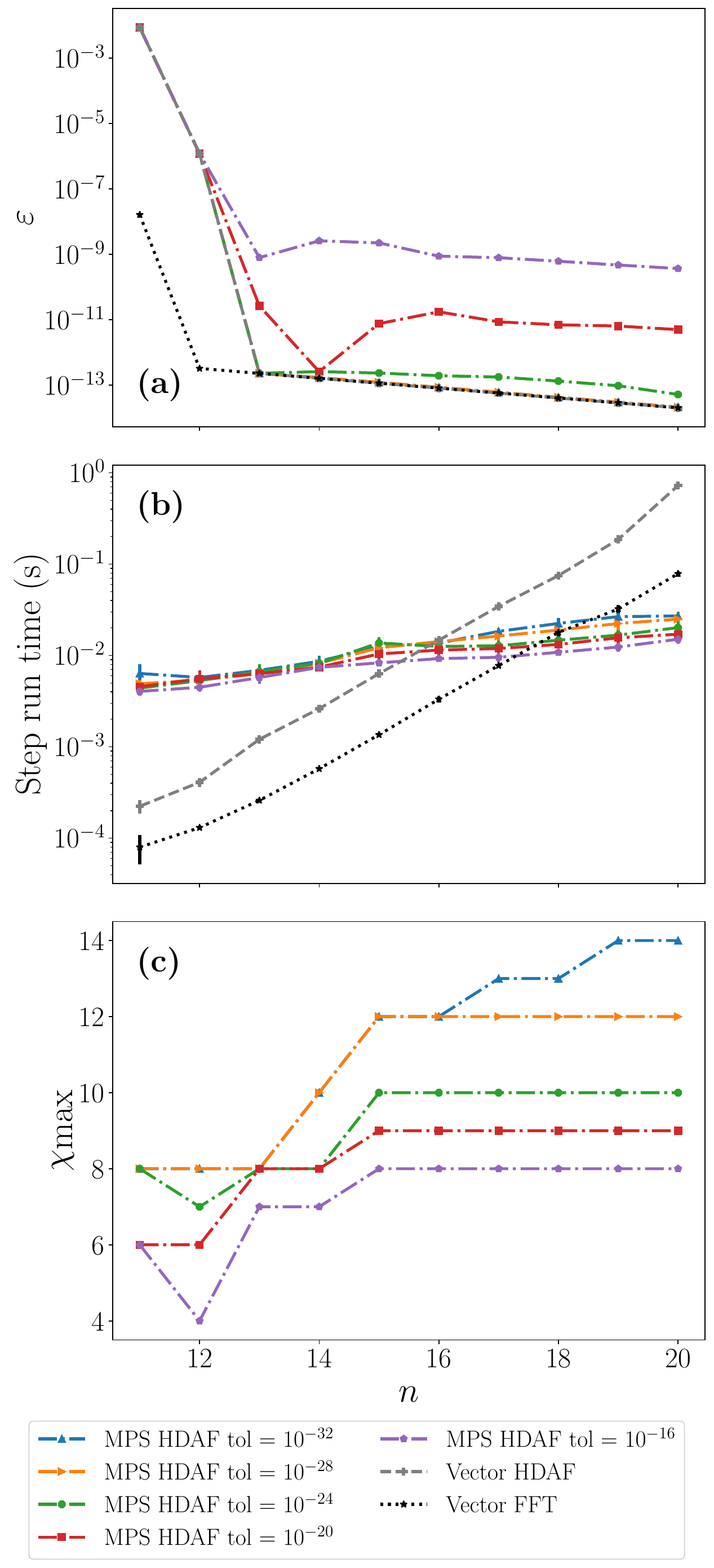}
  \caption{Number of qubits $n$ scaling of the split-step one-step evolution for vector-based---HDAF and FFT--- and different tolerance MPS-based HDAF for $\Delta t=0.0001$. (a) Error $\varepsilon$. (b) Run time. (c) Maximum bond dimension $\chi_\text{max}$ . The run time is averaged over ten runs.}
  \label{fig:mps_vs_vector_n}
\end{figure}

\begin{figure}
  \centering
  \includegraphics[width=\linewidth]{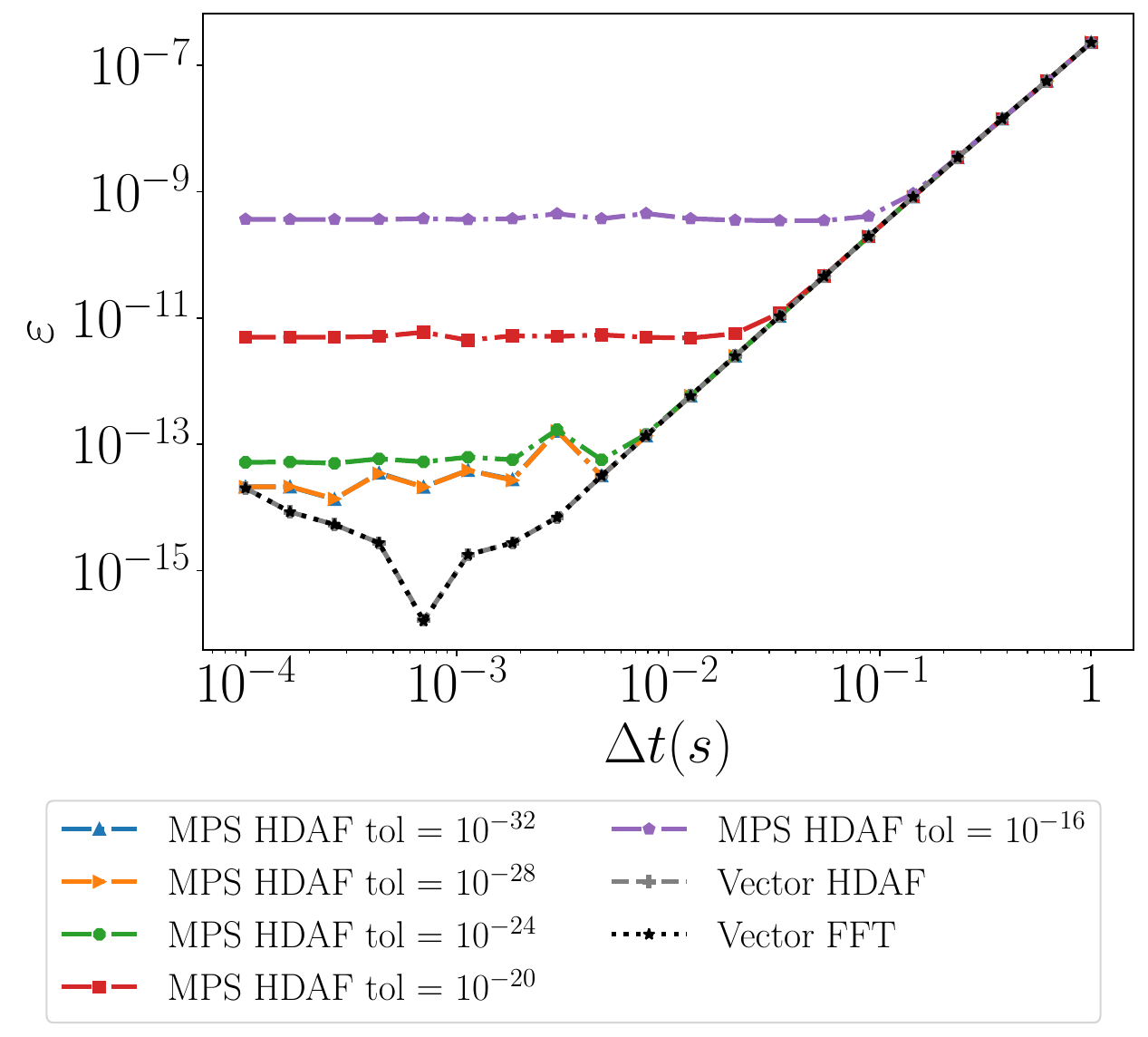}
  \caption{Error $\varepsilon$ scaling with time step $\Delta t$ for the split-step one-step evolution for vector-based---HDAF and FFT--- and different tolerance MPS-based HDAF for $n=20$.}
  \label{fig:mps_vs_vector_dt}
\end{figure}

All methods described before must be iteratively applied over small intervals $\Delta{t}$ to compose the whole evolution of a quantum state $\psi(x,t)$. A study and comparison of all methods over one such integration step is a good proxy for understanding the performance and accuracy of the algorithms over longer simulations.

The benchmark problem in this study will be the evolution of a quenched state under the Hamiltonian~\eqref{eq:ho_potential}. This analytically solvable problem can be used to estimate the errors in the wavefunction for different algorithms, grids, and time steps. Our simulation studies the expansion of a quantum state as we reduce the trapping frequency of the harmonic potential by a factor $\omega_H/\omega_0=0.01$. This leads to a 100-fold quantum state expansion, increasing its standard deviation from $\sigma_0$ to $\sigma_\text{max}=100\sigma_{0}$. The simulation domain in position space is designed to capture the wavepacket at its maximum expansion. This means we will represent functions over $x \in [-L/2, L/2)$ with $L = 16\sigma_\text{max}$. Since the initial state is derived from a tightly confined potential, the initial wavefunction is narrowly concentrated around $x=0$, which sets a lower bound for the grid discretization and number of qubits to accurately represent the initial and final state. Note that loading a function that is mostly zero outside a narrow interval is a tricky task for the MPS, requiring us to pad the function with zeros to compensate for the possible sampling or representation errors of TT-Cross or Chebyshev methods on the tails of the exponential.

The accuracy and performance of the algorithms are gauged using three figures of merit: (i) the function norm-2 difference 
\begin{equation}\label{eq:error}
  \varepsilon =\sqrt{\sum_i |\psi(x_i,\Delta t)-\tilde{\psi}(x_i,\Delta t)|^2\Delta x},
\end{equation}
measures the accuracy of the methods---$\psi(x_i,\Delta t)$ is the analytic solution for $t=\Delta t$, and $\tilde{\psi}(x_i,\Delta t)$ is the one-step state approximated by the method---; and (ii) the run time and (iii) maximum bond dimension $\chi_\text{max}$ determine their cost. 


The one-step study must separately study the time and spatial discretizations to independently understand the performance of the numerical integration and the PDE representation methods. Regarding spatial discretization, section~\ref{sec:representation} already demonstrated the exponential advantage of HDAF methods compared to finite differences in approximating the derivatives. With this idea in mind, the first study uses a grid with $n=18$ qubits (262144 points) to explore the different integration techniques. The finite difference approximation uses filter nine in Ref.~\cite{Holoborodko2008} to enhance noise suppression, and we fix a value $M=40$ for the HDAF approximation. Both techniques use periodic boundary conditions.

Figure~\ref{fig:FD_vs_HDAF} shows the error $\varepsilon$~\eqref{eq:error} and the run time scaling with $\Delta t$ for finite difference and HDAF derivative approximation for all methods. The scaling has saturated to the theoretical one for each time evolution method using HDAF approximation (Figure~\ref{fig:FD_vs_HDAF}(b)), achieving numerical precision only limited by the MPS truncation. However, the truncation error of the finite difference approximation limits the accuracy of the evolution in Figure~\ref{fig:FD_vs_HDAF}(a), leading to an error above the plateau of the HDAF implementation. Appendix \ref{app:error} shows the exact numerical scaling of $\varepsilon$ with $\Delta t$ for these numerical simulations.

As shown in Figures~\ref{fig:FD_vs_HDAF}(c)-(d), the implementation cost of HDAF and finite-difference algorithms is similar. The explanation for this is that the cost is dominated by the bond dimensions of the MPS states, which are similar for both algorithms. Given this and the exponentially improved accuracy of HDAF operators, we will, from now on, focus on the HDAF spectral methods and abandon the finite difference approximations entirely.

All evolution algorithms involve a trade-off between accuracy---i.e., the approximation order---and the cost of the implementation, which Figures~\ref{fig:FD_vs_HDAF}(b)-(d) accurately describe. In the HDAF implementation, all methods converge with the number of qubits in terms of the error $\varepsilon$. The methods show an algebraic dependence of the error $\varepsilon$ with the step $\Delta t$ as predicted by the theory, except for small time steps, limited by the numerical accuracy of the MPS implementation. The Arnoldi $n_v=10$ method is the best-performing one in terms of accuracy since it enables the largest step sizes due to its higher order. The Runge-Kutta, Arnoldi $n_v=5$, and split-step methods reach similar accuracies for a considerable $\Delta t$ range. The smaller order of the Euler method and the conjugate gradient implementation in the Crank-Nicolson method limit the error obtained beyond the intrinsic MPS truncation error.

To identify a method with a suitable cost-accuracy trade-off, it is essential to analyze the error $\varepsilon$ in conjunction with the run time (Figure \ref{fig:FD_vs_HDAF}(d)). Despite the Arnoldi $ n_v=10$'s low error, its run time is two orders of magnitude slower than the split-step method. Consequently, even with smaller time steps $\Delta t$, the split-step method demonstrates a better balance of cost and accuracy. Additionally, practical implementations often do not require such high accuracy, allowing for larger time steps. Thus, the split-step method with the HDAF approximation of the propagator is the optimal choice for studying the expansion problem.

Let us compare the performance of MPS methods to their vector implementation using split-step methods and the state-of-the-art fast Fourier transform (FFT). Figure~\ref{fig:mps_vs_vector_n} analyzes the scaling of split-step MPS and vector-based implementations with the discretization size. The MPS method examines various values of the truncation tolerance for the SVD and the simplifications of the finite precision algebra, measured as the norm-2 difference $\left\Vert\psi-\phi\right\Vert^2$ between the original state $|\psi\rangle$ with bond dimension $\chi_\psi$ and the one projected in the subspace of MPS $|\phi\rangle$ with bond dimension $\chi_\phi$, $\mathrm{MPS}_{\chi_\phi}$, such that $\chi_\phi<\chi_\psi$.

\begin{figure*}[t!]
  \centering
  \includegraphics[width=0.9\linewidth]{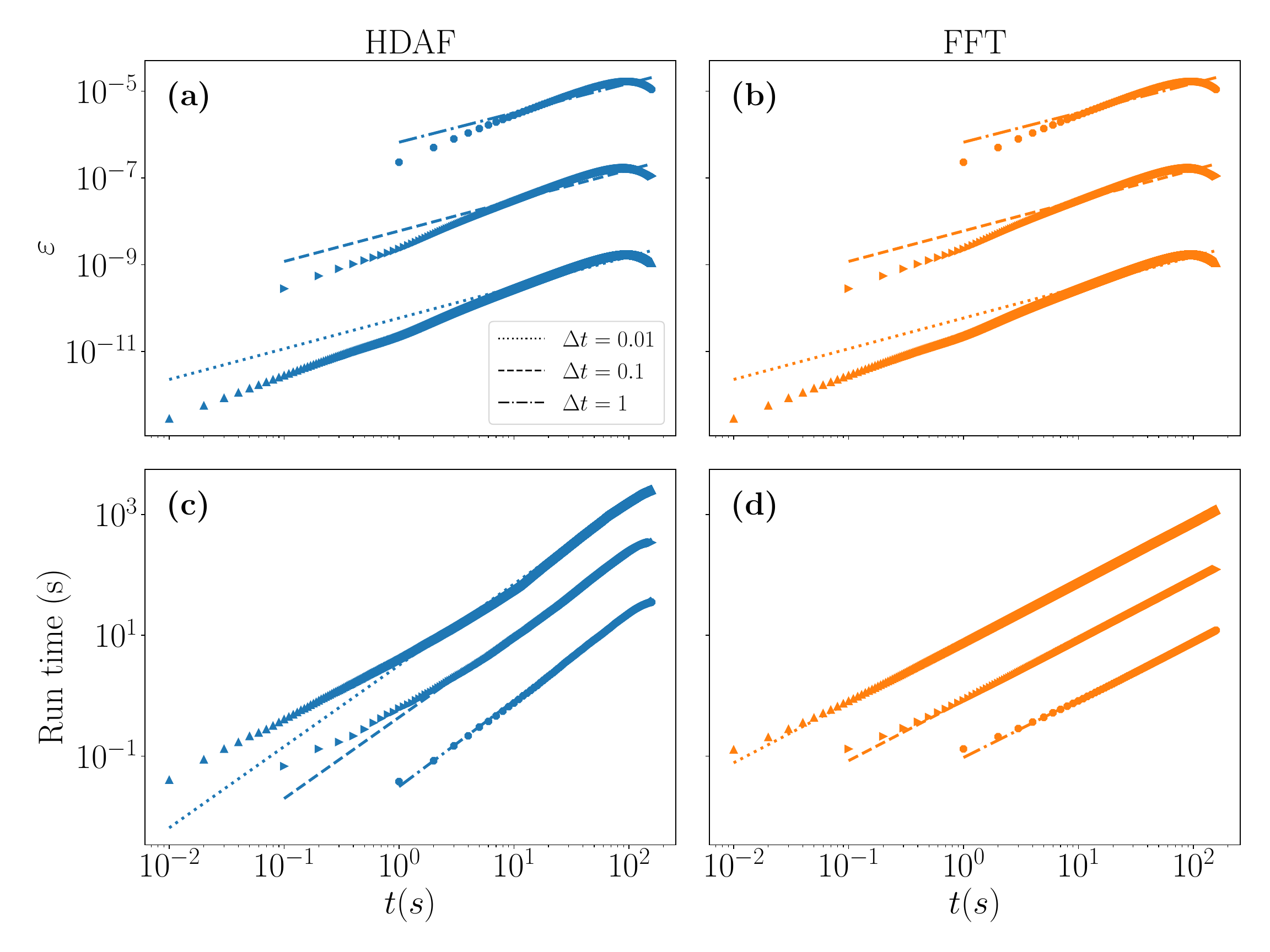}
  \caption{Particle expansion in a harmonic potential~\eqref{eq:ho_potential} with $\omega_H/\omega_0=10^{-2}$, $t_f=158$ and $n=20$ for the MPS split-step methods with HDAF differentiation and a state-of-the-art FFT split-step. Figures (a)-(c) show the error scaling with the evolution time, and figures (d)-(f) show the run time scaling with the evolution time.
  }\label{fig:evolution}
\end{figure*}

Figure~\ref{fig:mps_vs_vector_n}(a) demonstrates that the MPS tolerance dominates the split-step error. Achieving numerical precision comparable to the vector implementation requires tolerances of order $O(10^{-28})$ or smaller. This error scaling with the number of qubits demonstrates that a discretization with $\Delta x \approx 10^{-1}$ suffices for HDAF to converge well above the finite difference approach. Regarding the run time (Figure~\ref{fig:mps_vs_vector_n})(b)), the FFT is more efficient than the vector-based HDAF, and MPS perform asymptotically better than vector approaches, as they present an exponential scaling of time with the number of qubits due to the exponential increase in the number of points. The run time of the MPS method is similar regardless of the tolerance since all resulting states achieve a bond dimension of a similar order. Since tolerances below $10^{-28}$ overestimate the bond dimension needed, we choose this value for the SVD truncation in previous and future simulations while keeping a tolerance of the order of the numerical precision for MPS simplification. Finally, Figure~\ref{fig:mps_vs_vector_dt} shows that the MPS error $\varepsilon$ scaling with the time step $\Delta t$ presents a similar behavior to the vector-based implementation for tolerances smaller than $10^{-28}$. As the tolerance increases, it limits the accuracy for smaller time steps, for which the MPS truncation error dominates over the split-step truncation error. As the time step increases, the error associated with the method allows for larger tolerances.

\section{Quantum quench evolution}\label{sec:application}

After discussing the MPS-based algorithms for solving time-dependent PDEs, this section demonstrates the utility of the methods in a physical application. The problem under study has been presented in Section~\ref{sec:problem} and consists of the expansion of a quantum particle in a broad potential, which may be harmonic, as in Section\ref{sec:one-step}, or present a double-well structure (Section~\ref{sec:dw}). The study will focus on the algorithms that show the best balance between performance and accuracy, namely the split-step methods implemented with a fast-Fourier transform in the vector case and with HDAF in the MPS/QTT simulations.

\subsection{Harmonic expansion}\label{sec:ho}

The first set of simulations addresses a similar problem to Section~\ref{sec:one-step}, studying the expansion of a particle in a harmonic potential~\eqref{eq:ho_potential} that is 100 times weaker than the trap that confines the initial wavepacket. The simulation results in a 100-fold expansion of the wavefunction, which is properly captured with a discretization using $2^{20} = 1048576$ points or $n=20$ qubits for $x\in[L/2, L/2)$ with $L=16\sigma_\text{max}$. The expansion starts with the state of the original harmonic oscillator with frequency $\omega_0=1$, which is a real-valued Gaussian function. The potential is instantaneously weakened, relaxing the trapping frequency to $\omega_H=0.01$. This results in an acceleration of the wavepacket, an expansion that lasts until $t_f=0.5\pi /\omega_H$, at which the Gaussian solution~\eqref{eq:analytic_solution} reaches its maximum width. Let us consider three values for the time steps $\Delta t = 0.01,0.1,1$ and a final time $t_f=158$, so it fits all time steps. As in Section~\ref{sec:one-step}, $x\in[L/2,L/2)$ with $L=16\sigma_\text{max}$.

Figures~\ref{fig:evolution}(a)-(b) present the scaling of the error $\varepsilon$~\eqref{eq:error} with the time $t$ of the evolution for the MPS HDAF and vector FFT implementations of the split-step method, respectively. Note how the errors follow algebraic laws with very similar coefficients (see Appendix~\ref{app:evol}) that we attribute to the Störmer-Verlet integration scheme used: The error growth in time is consistent with the linear accumulation expected from a symplectic integration algorithm on a periodic Hamiltonian system~\cite{Cano_Sanz-Serna_1997}, as opposed to the quadratic law expected from general non-symplectic integrators such as Runge-Kutta on the same scenario~\cite{Stuart1998-ww}. The saturation of the global error could be explained since symplectic integrators keep the energy errors bounded at all times~\cite{Yoshida_1992}, also implying a bound on phase-space errors for conservative systems with periodic orbits. These properties make the split-step method very suitable for long-time simulations.

The run time scaling (Figures~\ref{fig:evolution}(c)-(d)) is close to linear for the FFT algorithms, as the problem size fixes the cost of each step, and the total run time is the sum of individual evolution steps. In contrast, the run time of the MPS simulation is slightly above linear, a fact that can be explained by the growth of the bond dimension as the wavefunction expands, leading to an increase in memory size and also in the cost of various MPS and MPO-MPS operations.

\begin{figure}[t]
  \centering
  \includegraphics[width=1\linewidth]{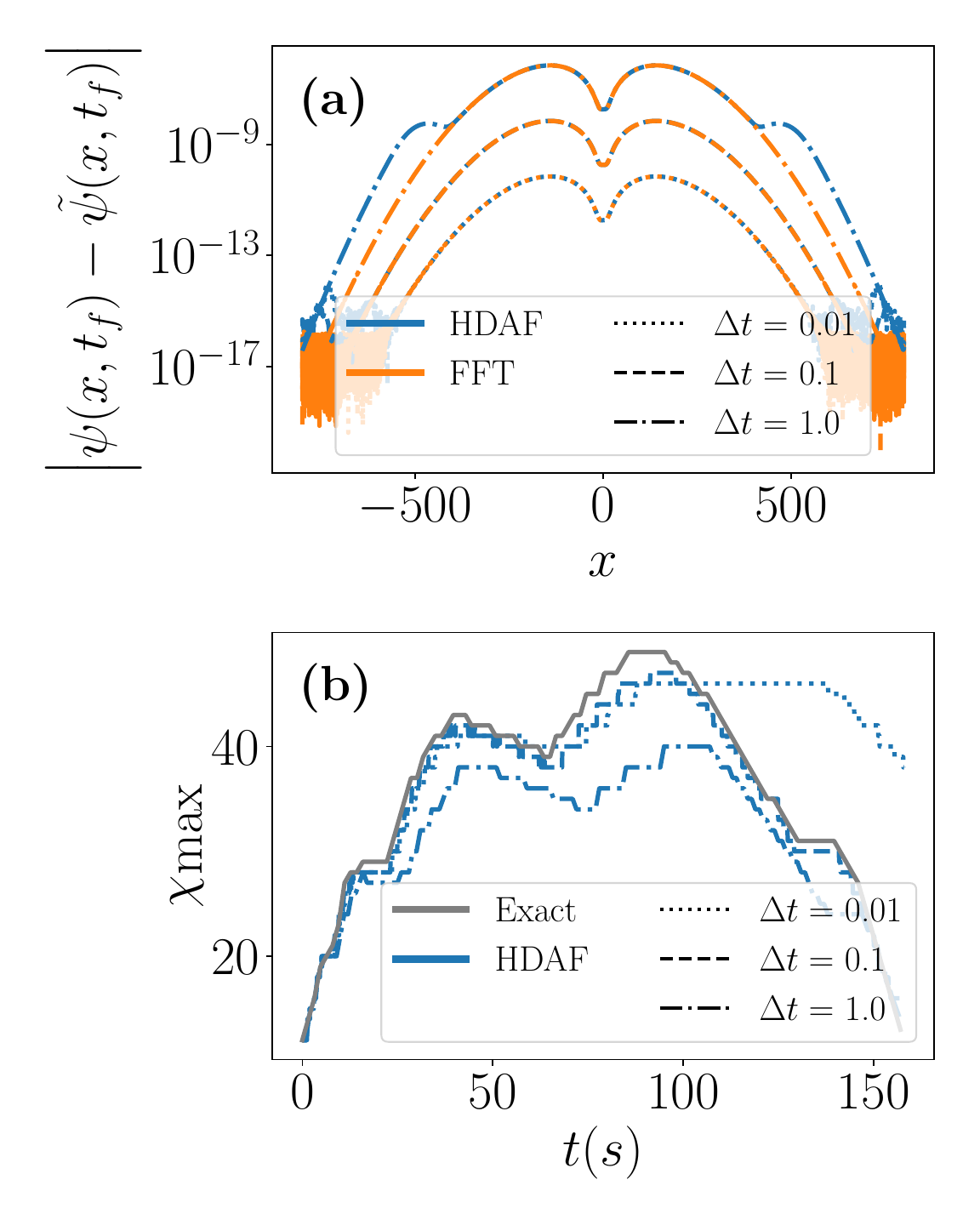}
  \caption{Particle expansion in a harmonic potential~\eqref{eq:ho_potential} with $\omega_H/\omega_0=10^{-2}$, $t_f=158$ and $n=20$ for the split-step methods with HDAF differentiation and a state-of-the-art FFT split-step. (a) Pointwise error $|\psi(x,t)-\tilde{\psi}(x,t)|$ of the maximum width solution solution approximated by the methods $\tilde{\psi}(x,t)$ with respect to the analytic solution $\psi(x,t)$~\eqref{eq:analytic_solution}. (b) Maximum bond dimension $\chi_\text{max}$ for each time step.}
  \label{fig:evolution_sol}
\end{figure}

The MPS algorithms' run time depends on the state's bond dimension at each time. Figure~\ref{fig:evolution_sol}(b) shows the maximum bond dimension $\chi_\text{max}$ for each time step. The numerical methods exhibit a similar behavior for the bond dimension of the solution in time, compared to the exact evolution, i.e., the bond dimension increases with the absolute value of the phase $\beta(t)$ of the analytic solution~\eqref{eq:analytic_solution}, with the lower bond dimensions associated to the initial and final states, which are real Gaussians. This behavior appears due to the chirping of the wavefunction, which is inherent to the physical setting and does not depend on the numerical method used, up to some precision allowance.
However, the bond dimension of the exact solution acted mainly as an upper bound to the bond dimension of the evolution we found, suggesting that the errors induced by our numerical methods deviated the solution not in a random direction but towards a more efficient MPS representation. This is consistent with the HDAF theory, where, before discretization, the approximations arise solely due to the attenuation of high frequencies. The HDAF operators seem to be well-suited to the MPS/QTT framework by relying on approximations that are a good fit for their formalism.

Figure~\ref{fig:evolution_sol}(a) shows the pointwise error of the maximum width state for the particle expansion, ${|\psi(x,t)-\tilde{\psi}(x,t)}|$, where ${\tilde{\psi}(x,t)}$ is the solution approximated by the numerical methods and  $\psi(x,t)$ is the analytic solution~\eqref{eq:analytic_solution}. We observe that both implementations of split-step have similar error shapes that differ on the extremes of the interval for the larger step size, possibly due to errors associated with the MPS representation.

\begin{figure}[t]
  \centering
  \includegraphics[width=1\linewidth]{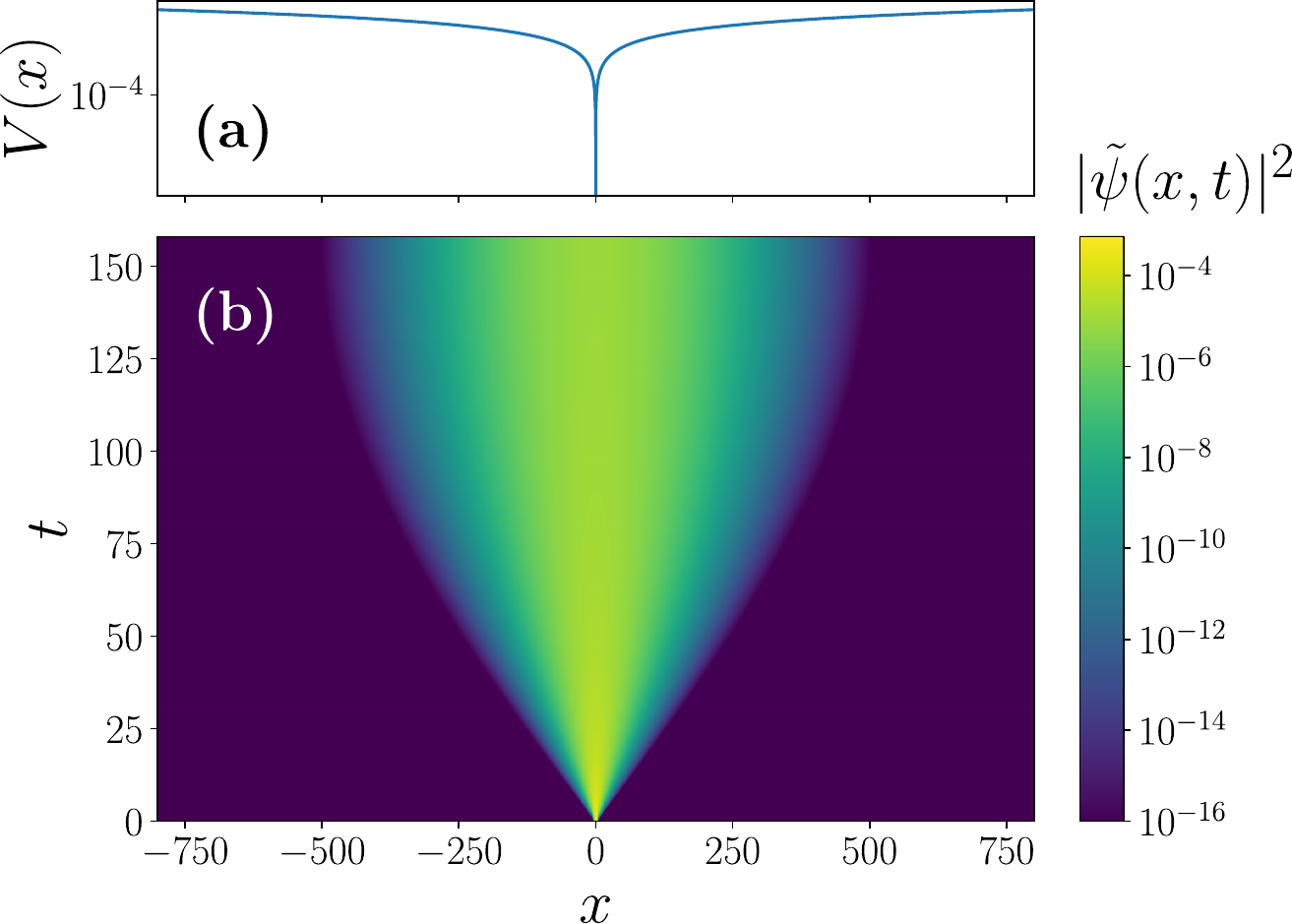}
  \caption{Particle expansion in a harmonic oscillator potential~\eqref{eq:ho_potential} with $\omega_H/\omega_0=10^{-2}$, with $t_f=158$, $\Delta t=0.1$ and $n=20$ for the MPS split-step method with HDAF differentiation. (a) Potential $V(x)$~\ref{eq:ho_potential}. (b) Wavefunction density $|\tilde{\psi}(x,t)|^2$.}
  \label{fig:ho}
\end{figure}

Let us focus on a concrete case to study the evolution of the wavepacket. Figure~\ref{fig:ho} depicts the evolution computed using the MPS HDAF split-step method for $\Delta t = 0.1$. As predicted by the analytic solution~\eqref{eq:analytic_solution}, the harmonic potential (Figure~\ref{fig:ho}(a)) induces an expansion of the particle, which is depicted in Figure~\ref{fig:ho}(b). 

\subsection{Double well potential}~\label{sec:dw}

\begin{figure}[t]
  \centering
  \includegraphics[width=1\linewidth]{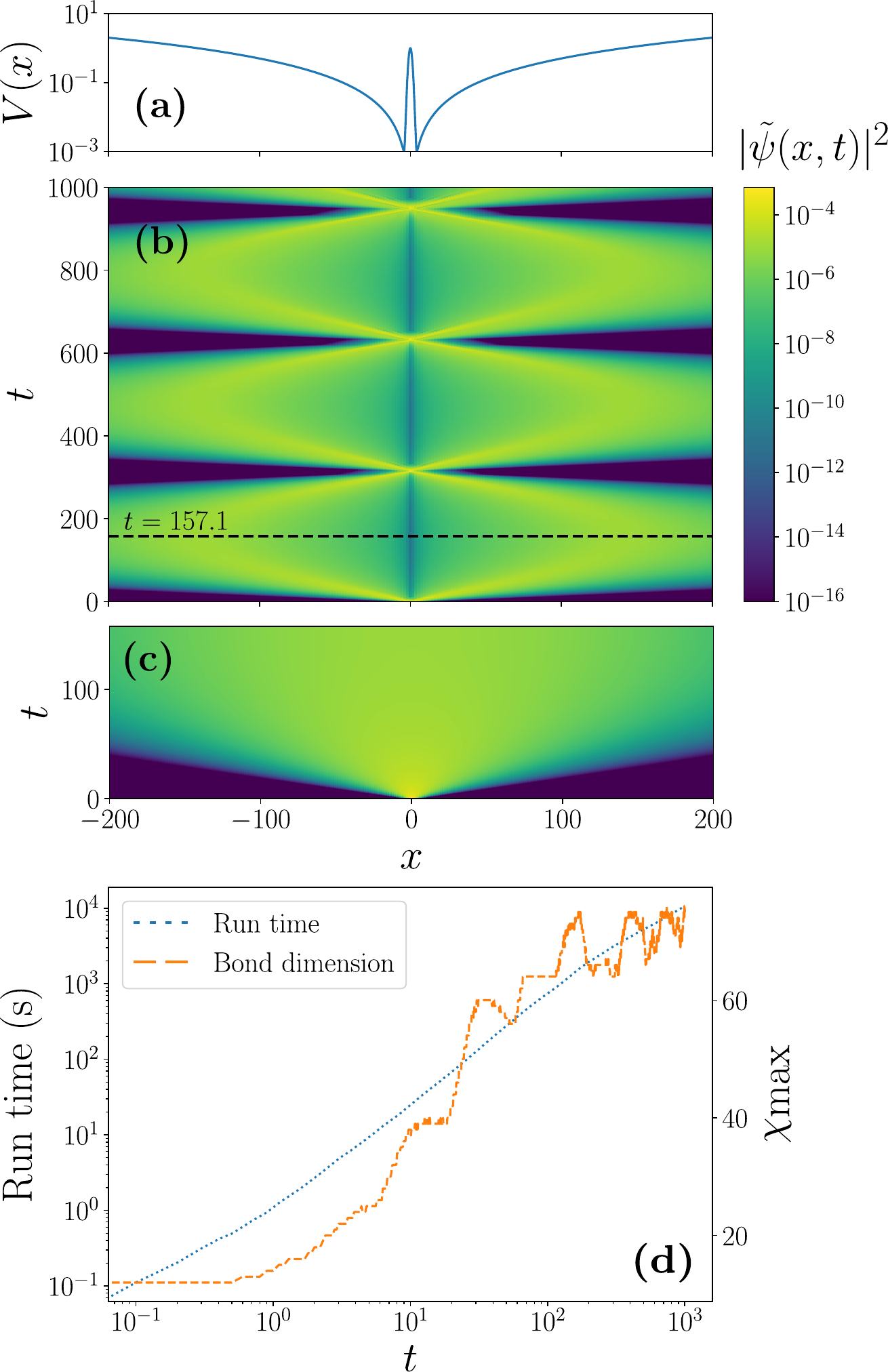}
  \caption{Particle expansion in a double well potential~\eqref{eq:double_well} with $\omega_H/\omega_0=10^{-2}$, $u=1$ and $\sigma=1$, with $t_f=1000$, $\Delta t=0.1$ and $n=20$ for the MPS split-step method with HDAF differentiation. (a) Potential $V(x)$~\ref{eq:double_well}. (b) Wavefunction density $|\tilde{\psi}(x,t)|^2$. (c) Wavefunction density $|\tilde{\psi}(x,t)|^2$ for the harmonic potential~\eqref{eq:ho_potential}. (d) Run time and maximum bond dimension $\chi_\text{max}$.}
  \label{fig:double_well}
\end{figure}

The calibration of simulation conditions from section~\ref{sec:ho} allows us to make informed decisions on the split-step algorithm, spatial discretization, time steps, and truncation errors. Let us now use this information to discuss a more interesting problem: the expansion of a nanoparticle in an anharmonic double-well potential. This problem is equivalent to a ``double-slit'' experiment for the particle, which will ideally spread into a coherent superposition of both halves of the trapping potential, and it constitutes a similar application to other problems studied in levitodynamics~\cite{Levitodynamics, Setter2019, Neumeier2024, RodaLlordes2024b}. For this particular simulation, the double-well potential is an open harmonic trap divided by a small Gaussian perturbation~\eqref{eq:double_well}, using $u=1$ and $\sigma=1$, and a trapping frequency that is once more 100 times smaller than the initial particle trap. The potential is depicted in Figure~\ref{fig:double_well}(a). As in the previous case, the harmonic term has a frequency $\omega_H<\omega_0$, which weakens the confinement and expands the particle. The Gaussian term is repulsive since $u>0$, separating the larger potential in two wells, with a barrier around $x=0$.

For this simulation, the final expansion time chosen is $t_f=1000$, which surpasses one period of the harmonic quantum quench solution $T=\pi/\omega_H$. This larger number is designed to explore multiple cycles of expansion and contraction of the particle's wavefunction to reveal collapse and revival dynamics. Figure~\ref{fig:double_well} shows simulation results, both from the particle's dynamics perspective and the wavefunction's complexity. During the evolution, the wavefunction's density $|\tilde{\psi}(x,t)|^2$ (Figure~\ref{fig:double_well}(b)) evolves according to the interplay of both terms in the potential~\eqref{eq:double_well}. The Gaussian term induces a separation in the particle's probability density. In contrast, the harmonic term preserves the confinement. It determines the period of the evolution, where the time of the maximum harmonic expansion $t_f=0.5\pi/\omega_H\approx 157.1$ coincides with the maximum spread of the particle. The new term modifies the behavior of the harmonic potential~\eqref{eq:ho_potential}, which is depicted in Figure~\ref{fig:double_well}(c). As expected, the harmonic potential induces cyclic expansion of the wavepacket, and the added Gaussian term leads to a barrier in the potential that divides the probability density into two localization peaks traveling in opposite directions.
The Gaussian barrier also modifies the behavior of the state's bond dimension. It is no longer cyclic, like in the harmonic case (Figure~\ref{fig:double_well}(d)), but instead seems to saturate, leading to a decrease in the linear scaling of the run time as the system evolves. 

\section{Conclusions}\label{sec:conclusions}

This work has introduced an HDAF encoding of differential operators for PDEs in an MPS/QTT framework. This encoding has shown exponential accuracy and low resource scaling. 

The MPS HDAF encoding enables the design of quantum-inspired time evolution algorithms to solve time-dependent PDEs: explicit and implicit Runge-Kutta methods, restarted Arnoldi iteration, and a split-step method. In particular, the split-step method benefits from the approximate representation of the free propagator unitary operator, which standard finite difference schemes cannot efficiently approximate.

The time evolution methods combined with HDAF overcome their finite difference implementation in terms of accuracy while maintaining a similar cost. Additionally, the split-step method shows the best trade-off in accuracy and cost. The HDAF time evolution algorithms are also competitive with state-of-the-art vector representations, enabling exponentially efficient encodings of functions and moderate overheads in the simulation.

The expansion of a particle in a broad potential acts as a benchmark for the methods. This poses a challenging problem since it defies the MPS representation due to the appearance of a chirp and rapid oscillations in the phase. Despite this chirp, the MPO-MPS algorithm produces accurate results with moderate bond dimensions and adequate run time.

Note that while the MPS and FFT speeds are comparable, only the MPS algorithm can scale up these grid densities to more dimensions. We expect to further develop and optimize these routines for higher-dimensional problems, accelerating them through C/C++ backends and other low-level optimizations to improve our Python programs' run time prefactors.

The present implementation is based on the SElf-Explaining Matrix Product State (SeeMPS) library for Python~\footnote{\url{https://github.com/juanjosegarciaripoll/seemps2}}.

\section*{Acknowledgments}
The authors would like to thank Juan José Rodríguez-Aldavero for his help with implementing the TT-cross interpolation. This work has been supported by Spanish Projects No. PID2021-127968NB-I00 and No. PDC2022-133486-I00, funded by MCIN/AEI/10.13039/501100011033 and by the European Union “NextGenerationEU”/PRTR”1. This work was also supported by the Ministry for Digital Transformation and of Civil Service of the Spanish Government through the QUANTUM ENIA project call - Quantum Spain project, and by the European Union through the Recovery, Transformation and Resilience Plan - NextGenerationEU within the framework of the Digital Spain 2026 Agenda. PGM acknowledges the funding by “FSE invierte en tu futuro” through an FPU Grant FPU19/03590
and by MCIN/AEI/10.13039/501100011033. JJGR and PGM acknowledge support
from CSIC Interdisciplinary Thematic Platform (PTI) Quantum Technologies (PTIQTEP+).
JG was supported by the Chilean National Agency for Research and Development (ANID-Chile), program ``Doctorado Nacional 2020'', scholarship No. 21202616.
The authors also gratefully
acknowledge the Scientific Computing Area (AIC), SGAI-CSIC, for their assistance
while using the DRAGO Supercomputer to perform the simulations.

\section*{Author contribution statement}

JG developed the extension of HDAF to the MPS formalism, its theoretical and numerical study, and the code implementation. PGM introduced the HDAF operators in the time evolution schemes and conducted their numerical characterization in the one-step and long-time simulations. JJGR conceptualized this work and its research goals. JJGR and LT supervised the research, providing solutions to challenges encountered throughout the process. JG and PGM wrote the original draft, and JJGR reviewed and edited the manuscript.

\appendix

\section{One-step $\varepsilon$ scaling with $\Delta t$}\label{app:error}
Figures~\ref{fig:FD_vs_HDAF}(a)-(b) show a fit for the error $\varepsilon$ with $\Delta t$ for the methods in Section~\ref{sec:TimeEvolution}, for the finite difference and HDAF derivative approximation of the derivative, respectively. Tables~\ref{tab:error_fd} and~\ref{tab:error_hdaf} contain the concrete numerical data of the fit $\varepsilon=C\Delta t ^m$. We use a piecewise linear fit and show the data for larger $\Delta t$, which are not limited by the MPS accuracy.

\begin{table}[ht]
  \centering

  \begin{tabular}{|l|c|c|} 
    \hline
    Method                      & \(C\)          & \(m\)          \\\hline
    Euler                       & $2.29 \times 10^{-2}$ & $1.99$ \\\hline
    Improved Euler              & $3.37 \times 10^{-2}$ & $3.14$ \\\hline
    Runge-Kutta                 & $2.20 \times 10^{-2}$ & $4.18$ \\\hline
    Crank-Nicolson              & $8.87 \times 10^{-1}$   & $3.44$ \\\hline
    Arnoldi \(n_v=5\)          & $2.04 \times 10^{-3}$ & $3.79$ \\\hline
    Arnoldi \(n_v=10\)         & $7.88 \times 10^{-4}$ & $4.45$ \\\hline
  \end{tabular}
  \caption{Function error $\varepsilon$~\eqref{eq:error} fit, $\varepsilon = C\Delta t^{m}$, for each method for a $n=18$ discretization and finite difference approximation of the derivative.}
  \label{tab:error_fd}
\end{table}

\begin{table}[ht]
  \centering
  \begin{tabular}{|l|c|c|} 
    \hline
    Method                      & \(C\)          & \(m\)          \\\hline
    Euler                       & $2.13 \times 10^{-2}$ & $2.00$ \\\hline
    Improved Euler              & $2.88 \times 10^{-2}$ & $3.00$ \\\hline
    Runge-Kutta                 & $2.21 \times 10^{-2}$ & $4.97$ \\\hline
    Crank-Nicolson              & $2.52 \times 10^{-1}$   & $3.73$ \\\hline
    Arnoldi \(n_v=5\)          & $1.02 \times 10^{-3}$ & $4.92$ \\\hline
    Arnoldi \(n_v=10\)         & $1.16 \times 10^{-5}$ & $9.11$ \\\hline
    Split-step                  & $4.94 \times 10^{-7}$ & $2.96$ \\\hline
  \end{tabular}
  \caption{Function error $\varepsilon$~\eqref{eq:error} fit, $\varepsilon = C\Delta t^{m}$, for each method for a $n=18$ discretization and HDAF approximation of the derivative.}
  \label{tab:error_hdaf}
\end{table}

\section{Harmonic quantum quench evolution scaling}\label{app:evol}

Figure~\ref{fig:evolution} shows the error $\varepsilon$ scaling and run time of the harmonic quantum quench evolution with time. Tables~\ref{tab:evol1}-\ref{tab:evol4} contain the concrete numerical data of the fits $\varepsilon = Ct^m$ and $T = Ct^m$, where $T$ is the run time.

\begin{table}[ht]
  \centering
  \begin{tabular}{|c|c|c|} 
    \hline
    \(\Delta t\) & \(C\) & \(m\) \\\hline
    0.01 & $5.89 \times 10^{-11}$ & 0.71 \\\hline
    0.1 & $6.00 \times 10^{-9}$ & 0.70 \\\hline
    1.0 & $6.65 \times 10^{-7}$ & 0.68 \\\hline
  \end{tabular}
  \caption{Function error $\varepsilon$~\eqref{eq:error} fit, $\varepsilon = Ct^{m}$, for the split-step HDAF MPS method, using different step-sizes $\Delta t$.}
  \label{tab:evol1}
\end{table}

\begin{table}[ht]
  \centering
  \begin{tabular}{|c|c|c|} 
    \hline
    \(\Delta t\) & \(C\) & \(m\) \\\hline
    0.01 & $5.89 \times 10^{-11}$ & 0.71 \\\hline
    0.1 & $6.00 \times 10^{-9}$ & 0.70 \\\hline
    1.0 & $6.65 \times 10^{-7}$ & 0.68 \\\hline
  \end{tabular}
  \caption{Function error $\varepsilon$~\eqref{eq:error} fit, $\varepsilon = Ct^{m}$, for the split-step FFT method, using different step-sizes $\Delta t$.}
  \label{tab:evol2}
\end{table}

\begin{table}[ht]
  \centering
  \begin{tabular}{|c|c|c|} 
    \hline
    \(\Delta t\) & \(C\) & \(m\) \\\hline
    0.01 & 3.19 & 1.35 \\\hline
    0.1 & 0.44 & 1.34 \\\hline
    1.0 & 0.03 & 1.42 \\\hline
  \end{tabular}
  \caption{Run time fit, $T = Ct^{m}$, for the split-step HDAF MPS method, using different step-sizes $\Delta t$.}
  \label{tab:evol3}
\end{table}

\begin{table}[ht]
  \centering
  \begin{tabular}{|c|c|c|} 
    \hline
    \(\Delta t\) & \(C\) & \(m\) \\\hline
    0.01 & 7.66 & 1.00 \\\hline
    0.1 & 0.82 & 0.99 \\\hline
    1.0 & 0.09 & 0.95 \\\hline
  \end{tabular}
  \caption{Run time fit, $T = Ct^{m}$, for the split-step FFT method, using different step-sizes $\Delta t$.
  }
  \label{tab:evol4}
\end{table}

\bibliography{my_bibliography}

\end{document}